\documentclass[preprint,11pt]{article}
\usepackage[a4paper,
            bindingoffset=0.2in,
            left=0.75in,
            right=0.75in,
            top=1in,
            bottom=1in,
            footskip=.25in]{geometry}
\usepackage{amsmath, amsfonts, amssymb}
\usepackage{graphicx}
\usepackage{subcaption}
\usepackage{fancyhdr}
\usepackage{cite}

\usepackage[colorlinks=true, linkcolor=blue, citecolor=blue, urlcolor=blue]{hyperref}

\newcommand{\be}{\begin{equation}}
\newcommand{\ee}{\end{equation}}
\newcommand{\bea}{\begin{eqnarray}}
\newcommand{\eea}{\end{eqnarray}}

\begin{document}

\title{Complex dynamics and route to quasiperiodic synchronization in non-isochronous directed Stuart-Landau triads}

\author{
Ankan Pandey$^{1}$\footnote{ankan0506@gmail.com (Corresponding Author)}, 
Sandip Saha$^{2}$\footnote{sahasandip.loknath@gmail.com}, 
Dibakar Ghosh$^{3}$\footnote{dghosh.nld@gmail.com} \\[0.5em]
{\small $^1$Department of Chemical Sciences, Indian Institute of Science Education and Research, Berhampur, India} \\
{\small $^2$National Center for Biological Sciences, Tata Institute of Fundamental Research, Bangalore, India} \\
{\small $^3$Physics and Applied Mathematics Unit, Indian Statistical Institute, Kolkata, India}
}

\date{} 

\maketitle

\begin{abstract}
The coupled Stuart–Landau equation serves as a fundamental model for exploring synchronization and emergent behavior in complex dynamical systems. However, understanding its dynamics from a comprehensive nonlinear perspective remains challenging due to the multifaceted influence of coupling topology, interaction strength, and oscillator frequency detuning. Despite extensive theoretical investigations over the decades, numerous aspects remain unexplored, particularly those that bridge theoretical predictions with experimental observations—an essential step toward deepening our understanding of real-world dynamical phenomena. This work investigates the complex dynamics of unidirectionally coupled \textit{non-isochronous} Stuart-Landau oscillators. Calculations of steady-states and their stability analysis further reveal that periodic attractors corresponding to weak forcing or coupling regimes are dynamically unstable, which pushes the system towards quasiperiodic oscillation on the torus attractor. The mapping of parameter values with the kind of attractor of the oscillatory system is presented and classified into periodic, quasiperiodic, partially synchronized, and chaotic regions. The results of this study can be leveraged to design complex yet controllable dynamical architectures.
\end{abstract}


\section{Introduction}
\label{Intro}

The Stuart–Landau (SL) equation~\cite{landau1944problem, stuart1958non, stuart1960non} is one of the most fundamental mathematical models used to describe real-world dynamical systems across diverse domains such as physics, biology, and chemistry. This equation captures a wide range of nonlinear behaviors through its various forms and extensions~\cite{stuart1960non, kuramoto2003chemical}. Over the past few decades, it has been extensively studied to explore complex dynamical phenomena such as amplitude death~\cite{zou2014emergence, roopnarain2021amplitude}, transitions between steady states~\cite{stuart1958non,mendola2025collective,verma2025explosive}, synchronization~\cite{kuramoto2003chemical,bayani2023explosive}, coexistence of synchrony and asynchrony~\cite{abrams2004chimera,parastesh2021chimeras,majhi2019chimera}, and dynamical robustness \cite{majhi2024dynamical}.

To study such complex behaviors in real-world systems, a common approach is to incorporate multiple Stuart–Landau oscillators coupled to each other according to the system’s interaction rules~\cite{abrams2004chimera, garcia2012complex, lee2022nontrivial}. However, as the number of oscillators increases, the mathematical and numerical analysis becomes significantly more complex. Despite extensive research, the number of reported studies addressing large ensembles of coupled SL oscillators remains limited, primarily due to the analytical intractability of such systems~\cite{kemeth2019cluster, pikovsky2015dynamics}. Since there exists no general theoretical framework to handle large networks of SL oscillators analytically, numerical simulations often serve as the only feasible approach~\cite{duvsek1994numerical,pikovsky2015dynamics,kuramoto2003chemical,garcia2012complex}. However, numerical methods are inherently prone to discretization and approximation errors, which may obscure subtle dynamical features or lead to incomplete interpretations of system behavior.

Investigating these complex phenomena requires considering multiple Stuart–Landau oscillators that can interact through various coupling topologies~\cite{kumar2021two,zhao2018enhancing,haugland2023coexistence,li2022mean}. Both the coupling strength and network topology play crucial roles in determining the overall system dynamics~\cite{kumar2021two,pandey2020coupled,hashemi2025inter,serrano2025fractional,li2022mean,pham2019simulation,serrano2022robust,khennaoui2019chaos}. Variations in topology can lead to subtle yet important changes in emergent behaviors~\cite{kumar2021two,ghosh2022synchronized}. Given the vast space of possible coupling configurations, systematically studying all interaction schemes among even a finite set of SL oscillators presents significant challenges. Conducting such exhaustive investigations is both computationally expensive and analytically demanding, emphasizing the need for focused approaches to better understand their underlying mechanisms.

Among the different coupling schemes, unidirectional coupling~\cite{perlikowski2010routes,zhao2025nonreciprocal,barba2023dynamics,vijayan2025lag} represents a particularly simple yet intriguing form. In this setup, one oscillator (the source or driver) influences another (the target or response), but not vice versa. More specifically, each oscillator acts as a driver for the next, forming a directed cascade. This one-way interaction, in contrast to bidirectional or globally coupled networks, introduces a hierarchical flow of information within the system.  This structure allow investigation of how dynamical properties—such as frequency entrainment, amplitude modulation, and phase shifts—evolve along the chain~\cite{perlikowski2010routes,zhao2025nonreciprocal,barba2023dynamics,ryu2017amplitude}. As the number of oscillators increases, diverse dynamical regimes emerge, reminiscent of graded response patterns and dynamical filtering phenomena. Mathematically, unidirectional coupling can be implemented by adding a coupling term to the evolution equation of the target oscillator while keeping the source oscillator unaffected~\cite{perlikowski2010routes,zhao2025nonreciprocal,barba2023dynamics,ryu2017amplitude}. For example, in a system of two coupled SL oscillators $z_1$ and $z_2$, unidirectional coupling implies that $z_1$ evolves independently, while the dynamics of $z_2$ is driven by $z_1$. Such a coupling scheme is not merely a mathematical abstraction—it naturally appears in real-world systems such as neural signal propagation, unidirectional power grids, and laser arrays~\cite{perlikowski2010routes, zhao2025nonreciprocal, barba2023dynamics, ryu2017amplitude}.

From this general standpoint, unidirectional coupling provides a useful framework for understanding the global behavior of coupled Stuart–Landau oscillators through the study of driven–dissipative dynamics, phase locking, and entrainment phenomena in open systems. Prior studies have demonstrated that even in the absence of mutual feedback, complex synchronization patterns can emerge~\cite{pikovsky2001synchronization,nakao2016phase}. The resulting global dynamics may encompass diverse phenomena such as chaos, amplitude suppression, and the emergence of steady states~\cite{perlikowski2010routes,zhao2025nonreciprocal,barba2023dynamics,ryu2017amplitude}. However, identifying and characterizing these behaviors analytically remains challenging without extensive numerical simulations. In practical applications, unidirectional coupling is frequently utilized in master–slave architectures, feedforward neural networks, and delay-line oscillators, where precise control and predictability of system behavior are crucial~\cite{pyragas1992continuous}. Although several studies have reported such global behaviors, a systematic, step-by-step analysis of how these phenomena emerge locally through unidirectional (leader–follower) interactions remains largely unexplored.

Moreover, understanding the influence of intrinsic oscillator properties is equally important in shaping the collective dynamics of such unidirectionally coupled systems. In self-sustained oscillatory systems, isochronicity refers to the property that all trajectories approaching the limit cycle rotate with the same angular velocity, implying that the oscillation frequency is independent of amplitude~\cite{Calogero2008,limiso,calogero2011isochronous,saha2025power,saha2022existence}. However, many real-world oscillators deviate from this ideal behavior and exhibit non-isochronicity (or amplitude–phase coupling), where amplitude fluctuations induce shifts in the instantaneous frequency. The Stuart–Landau oscillator, as the canonical normal form of a Hopf bifurcation, provides a minimal framework to study such effects through the inclusion of a non-isochronicity parameter~\cite{kuramoto2003chemical,pikovsky2001synchronization}. Non-isochronicity introduces shear in the phase space, distorting isochrons and significantly influencing synchronization dynamics, phase response, and collective behavior in coupled oscillator networks~\cite{kuramoto2003chemical,pikovsky2001synchronization}.

In this work, we build upon earlier foundational studies on unidirectionally coupled oscillators~\cite{koseska2013oscillation,rosenblum2004controlling,reddy1998time}, extending the analysis to longer chains to explore the emergent dynamics as a function of coupling strength, heterogeneity, and initial conditions. We aim to address this gap by systematically investigating the local dynamical mechanisms underlying unidirectionally coupled Stuart–Landau oscillators in the presence of a non-isochronous term, which is commonly encountered in general SL systems. We demonstrate how complex global phenomena—such as the appearance of isolated branches (isolas) and routes to chaos—can be understood through successive leader–follower interactions. This approach not only provides a clearer mechanistic understanding of local-to-global transitions but also offers a potential framework to analyze and predict emergent behaviors in large networks of coupled oscillators. Our findings indicate that these systems exhibit modular and hierarchical dynamical organization, which may provide insights for designing controllable dynamical architectures—relevant for neuromorphic computing, signal processing, and synthetic biological networks.

The work is organized as follows: We start with two coupled SL oscillators in Sec.~\ref {CSLO} and develop necessary analytical expressions for further analysis. In Sec .~\ref {SteadyStateStudy}, we discuss long-time stationary solutions of the system, such as amplitude death, amplitude, and phase response, where a detailed mechanism of isola and resonance is provided. This is followed by Sec.~\ref{StabilityAnalysis} where qualitative stability analysis of the steady states is conducted. In Sec.~\ref{QuasiperiodOsci}, we report the existence and parametric dependence of quasiperiodic oscillations. In Sec.~\ref{TriadSLO}, a network of triadic SL network of oscillators is considered, and the response landscapes are calculated under quasiperiodic driving on the third oscillator. The detailed mechanism of quasiperiodic synchronization and chaos is also provided through Lyapunov exponent studies. The work summarised in Sec.~\ref{Summary}, where the discussion extends to possible applications and future work.

\section{Coupled Stuart-Landau oscillators}\label{CSLO}

From the extensive research on the SL oscillator, we focus on the simplest possible case to interpret the new results and observations obtained in our study. Considering more complex scenarios would make it significantly harder to extract and understand the fundamental behaviors. By restricting ourselves to this minimal configuration, we can gain clear insight into the underlying dynamics, which in turn helps us interpret more complex results. In this context, the unidirectionally coupled SL oscillator is described by:

\bea
\dot{Z_1} = (\lambda_1 + \iota \omega -(1 + \iota \alpha)|Z_1|^2)Z_1,  \\
\dot{Z_2} = (\lambda_2 + \iota \omega - (1 + \iota \alpha)|Z_2|^2)Z_2 +   \epsilon(Z_1 - Z_2). 
\label{eq:cSLOmain}
\eea

This is the normal form of Hopf bifurcation, which results in a limit cycle with radius $\lambda_1$, $\lambda_2$ of the two individual oscillators, respectively. The corresponding frequencies are $\omega_1$ and $\omega_2$. Parameter $\alpha$ is the non-isochronicity coefficient. The two LOs are dissipatively coupled with a coefficient $\epsilon$.  \\
For $Z_i = r_i e^{\iota \theta_i}$, the equations transforms to polar form

\bea
\dot{r_1} = (\lambda_1 - r_1^2)r_1, 
\label{eq:cSLOpolarr1} \\
\dot{r_2} = (\lambda_2 - \epsilon  - r_2^2)r_2 + 
\epsilon r_1 \cos(\theta_1 - \theta_2),
\label{eq:cSLOpolarr2}  \\
\dot{\theta_1} = \omega_1 - \alpha r_1^2,
\label{eq:cSLOpolartheta1}  \\
\dot{\theta_2} = \omega_2 - \alpha r_2^2 + \epsilon\frac{r_1}{r_2}\sin(\theta_1 - \theta_2). 
\label{eq:cSLOpolartheta2}
\eea

The variables $r_i = |Z_i|$ and $\theta_i$ represent the amplitude and phase of the $ith$ LO. Without coupling, both oscillators show a stable origin ($r=0$) for $\lambda_i <0$, which is called \textit{amplitude death} (AD) state. For $\lambda_i >0$, the individual oscillators show stable limit cycles with radius $\sqrt{\lambda_i}$. The individual phases are determined by the initial conditions. Different initial conditions determine at which point on the cycle the transient trajectory will approach. Without a non-isochronic term, oscillators will oscillate at frequency $\omega_i$. However, the non-isochronic term, $\alpha$, causes amplitude-dependent deflection from this frequency. Hence, oscillators with different amplitudes (= $\sqrt{\lambda}$) shall 
deviates differently from its intrinsic frequency.

\subsection{Analytical catalog of amplitude and phase response}
\label{CSLO-Analysis}

To proceed, let's consider the phase difference between the oscillators. As the two oscillators are asymmetric, it is prudent to define $p_1:p_2$ phase difference as $\phi = p_1\theta_1 - p_2\theta_2$, where $p_1, p_2$ represent the number of cycles completed by each oscillator in a given time. Using this definition, the difference $\theta_1 - \theta_2$ can be written as $\frac{p_2 - p_1}{p_2} \theta_1 + \frac{\phi}{p_2}$. The corresponding dynamics of $\phi$ are expressed as
\bea
\dot{\phi} = (p_1\omega_1 - p_2\omega_2) - \alpha (p_1 r_1^2 - p_2 r_2^2) - \epsilon\frac{r_1}{r_2} \sin\left(\frac{p_2 - p_1}{p_2} \theta_1 + \frac{\phi}{p_2}\right). 
\label{eq:phiEquMain}
\eea

Together with the amplitude equation, these equations determine the complete dynamics of the coupled oscillators. The fixed points of the leader amplitude oscillator dynamics are given as $r_1^*=0, \sqrt{\lambda_1}$, and the phase, $\theta_1(t) = (\omega_1-\alpha {r_1^*}^2)t = (\omega_1 - \alpha \lambda_1)t$.  
The fixed points of the dynamics of the follower oscillator can be obtained by solving 

\bea
(p_1\omega_1 - p_2\omega_2) - \alpha (p_1 r_1^2 - p_2 r_2^2) - \epsilon\frac{r_1}{r_2} \sin\left(\frac{p_2 - p_1}{p_2} \theta_1 + \frac{\phi}{p_2}\right) = 0, \\
(\lambda_2 - \epsilon  - r_2^2)r_2 + \epsilon r_1 \cos \left(\frac{p_2 - p_1}{p_2} \theta_1 + \frac{\phi}{p_2}\right) = 0.
\eea

After some algebraic manipulation, 

\bea
\left((p_1\omega_1 - p_2\omega_2)r_2 - \alpha (p_1 r_1^2 - p_2 r_2^2)r_2 \right)^2 + \left((\lambda_2 - \epsilon  - r_2^2)r_2 \right)^2 = \epsilon^2 r_1^2
\label{eq:amplitudeResp}
\eea

and

\bea
\phi = p_2 \arctan\left(\frac{-(p_1\omega_1 - p_2\omega_2)r_2 + \alpha (p_1 r_1^2 - p_2 r_2^2)}{\lambda_2 - \epsilon  - r_2^2}\right) - (p_2 - p_1)\theta_1.
\label{eq:phaseResp}
\eea

\section{Long-time stationary regimes} \label{SteadyStateStudy}

In this section, we will outline steady states found in the followers' dynamics. We will further explore the response of the follower's amplitude and phase on the leader's parameters.

\subsection{Amplitude death state}\label{SSS-AD}
The conditions for stability of the follower AD state can be obtained from its dynamics given in Eq.~\eqref{eq:cSLOpolarr2}.
However, it is not straightforward to obtain parametric conditions without some assumptions. However, a bound can be estimated based on the phase difference between the leader and the follower. To proceed, consider the linearisation of Eq.~\eqref{eq:cSLOpolarr2} around $r_2=0$, then the state is stable for 

\be
\dot{r_2} = (\lambda_2 - \epsilon)r_2 + \epsilon r_1 \cos(\theta_1 - \theta_2) < 0.
\ee

For an in-phase situation, $\cos(\theta_1 - \theta_2)=1$, Eq.~\eqref{eq:cSLOpolarr2} can be solved to give the following,

\bea
r_2(t) = -\epsilon\frac{r_1}{\lambda_2-\epsilon} + r_2(0) e^{(\lambda_2-\epsilon) t},
\eea 
where $r_2(0)$ is initial condition. Clearly, the AD state is not approachable even for the $\lambda_2 <0$ case. AD state is stable only when it is stable for the leader too, i.e., for $\lambda_1<0$. Further, for a stable leader AD, the follower AD is stable for $\lambda_2<\epsilon$. Above this threshold follower shall show periodic behavior.\\
For anti-phase, $\cos(\theta_1 - \theta_2)=-1$, the solution is given as
 
\bea
r_2(t) = \epsilon\frac{r_1}{\lambda_2-\epsilon} + r_2(0) e^{(\lambda_2-\epsilon) t}.
\eea  

In this case, again, AD is stable for $\lambda_2<0$ only if it is also a stable state of the leader. This result can be extended to non-synchronized cases as well. Therefore, for positive $\lambda_1, \lambda_2$ values, AD states are unstable and at least one stable periodic solution exists in the system. The time evolution approximation is near the origin. The trajectories will not blow with higher $\lambda_2$ values as they will be suppressed by the cubic nonlinearity in the amplitude dynamics. 

\begin{figure}[!h]
\centering
\includegraphics[width=1\linewidth]{"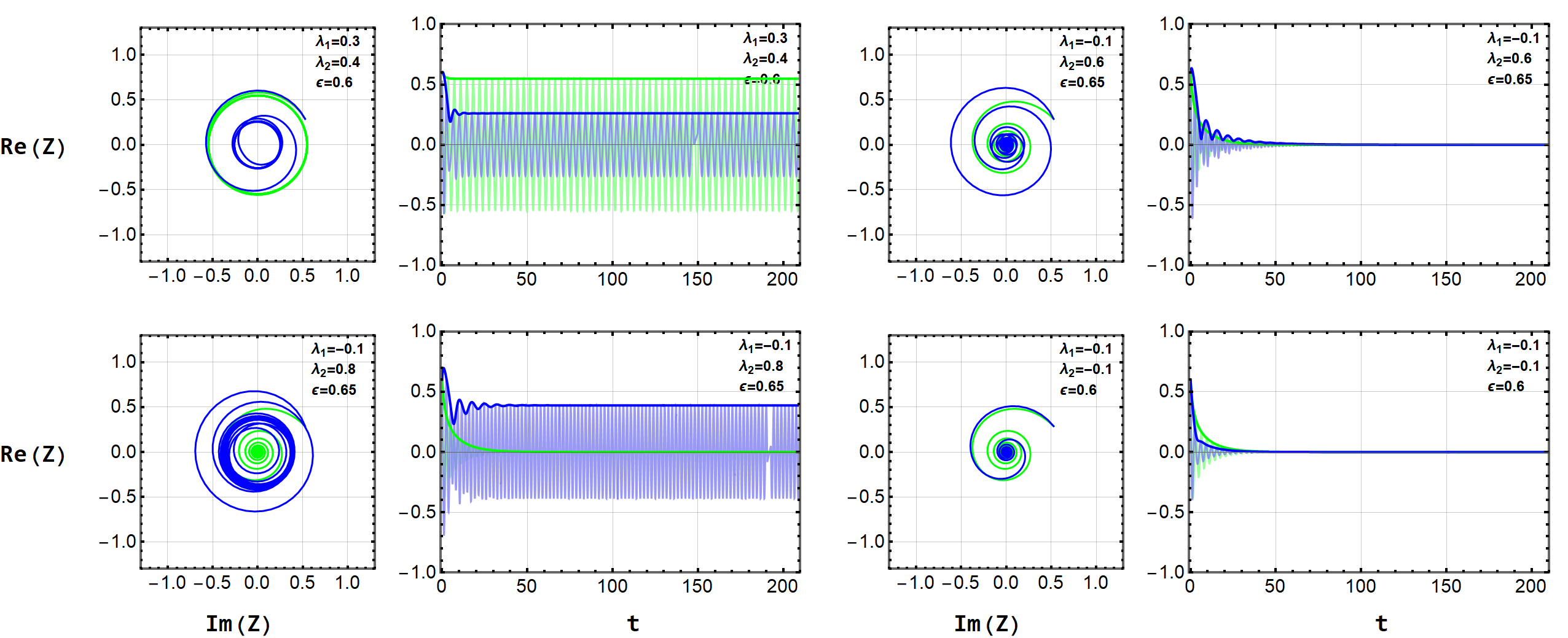"}
\caption{Illustration of amplitude death in the follower's dynamics. The plot depicts four scenarios with a phase portrait and a time series. The variable parameters are shown in the insets. Green lines correspond to the leader's dynamics and blue lines to the follower's dynamics. Parameters: $\alpha=(0.6,1.0,1.5)$, $r_2=1.0$, $\omega_2=3.0$. }

\label{fig:PPADPS}
\end{figure}

This behavior is shown in Fig.~\eqref{fig:PPADPS}. The plots are obtained by numerically solving Eq.~\eqref{eq:cSLOmain} with initial conditions $r_1(0)=0.6$, $r_2(0)=0.6$, $\theta_1(0)=0.5$, and $\theta_2(0)=0.5$. The static parameters are $\omega_1=2.0$, $\omega_2=3.0$, $\alpha=1.0$. The figure shows the phase portrait and time-series dynamics of the leader and follower oscillator and their variation with parameters $\lambda_1, \lambda_2, \epsilon$. In the top left block $\lambda_1, \lambda_2 >0$, and hence both the oscillators show periodic behavior. This is independent of the coupling coefficient. However, for $\lambda_1<0$ and $\lambda_2<\epsilon$, the amplitude decays to the AD state, shown in the top right block. As $\lambda_2 (=0.8)$ crosses the $\epsilon (=0.65)$ value, Hopf bifurcation leads to periodic solutions for the follower, even though the leader is in AD state. This is shown in the bottom left block. Finally, for $\lambda_1<0, \lambda_2<0$, AD becomes stable for both the oscillators and is depicted in the bottom right block.

\subsection{Phase response}\label{SSS-PhsRes}

For certain parameter conditions, the phase dynamics of the follower lock to the leader phase over time. For locking the dynamics of the generalized phase, $\phi = p_1\theta_1-p_2\theta_2$, must be bounded. To estimate the boundedness of the resonant phase average over the fast dynamics of the leader's phase $\theta_1 = (\omega_1-\alpha \lambda_1)t$ and consider the follower phase dynamics to be slow. The phase dynamics Eq.~\eqref{eq:phiEquMain} becomes
\be 
\dot{\phi} = (p_1\omega_1 - p_2\omega_2) - \alpha (p_1 r_1^2 - p_2 r_2^2) - \epsilon\frac{r_1}{r_2} A_{p_1:p_2} \sin\left(\frac{\phi}{p_2}\right),
\ee
where 
$$ A_{p_1:p_2} = \frac{p_2 \sin \left(\frac{\pi (p_2-p_1)(\omega_1 - \alpha\lambda_1)}{p_2}\right)}{\pi (p_2-p_1)(\omega_1 - \alpha\lambda_1)}.
$$

This essentially is the projection of the leader on the follower's phase. For $p_1=p_2$ case, $A_{p_1:p_2} = 1$. Now, for bounded dynamics, the detuning $(p_1\omega_1 - p_2\omega_2) - \alpha (p_1 r_1^2 - p_2 r_2^2)$ must be less than the effective coupling; therefore, the phase-lock conditions are given as
\be 
|(p_1\omega_1 - p_2\omega_2) - \alpha (p_1 r_1^2 - p_2 r_2^2)| <  \epsilon\frac{r_1}{r_2} A_{p_1:p_2}.
\label{eq:phsLockCond}
\ee
 
In regions formed by the parameter values which satisfy Eq.~\eqref{eq:phsLockCond}, the phase is locked and the solutions are $p_1:p_2$ synchronized with the leader oscillator. These regions in the parametric plane are referred to as \textit{Arnold tongues}. For values outside the \textit{tongues}, the phase drifts away and doesn't stay steady, resulting in quasi-periodic dynamics. 

\begin{figure}[]
\centering
\includegraphics[width=0.9\textwidth]{"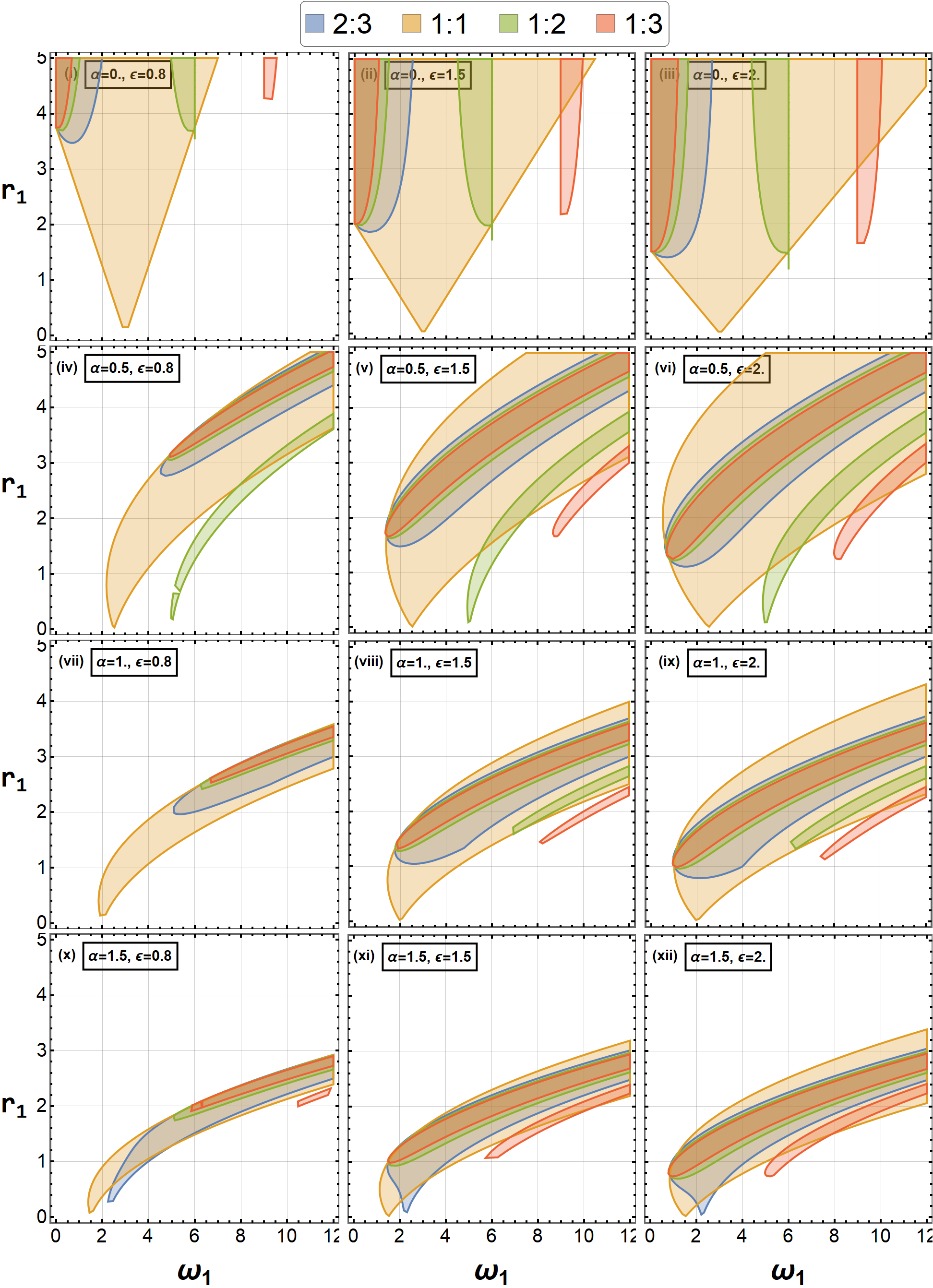"}

\caption{ \textit{Arnold Tongues} for \textbf{$(2:3$, $1:1$, $1:2$, $1:3)$} phase locking of follower oscillator in $r_1-\omega_1$ plane. From left to right coupling coefficient $\epsilon$ increases and plotted for $\epsilon=(0.8, 1.5, 2.0)$. From top to bottom non-isochronous parameter $\alpha$ increases and is plotted for $\alpha=(0.6,1.0,1.5)$. For the plot $r_2=1.0$, and $\omega_2=3.0$. }

\label{fig:ATsnaps}
\end{figure}

The qualitative nature of stability regions depends on the system parameters of both oscillators. We are primarily interested in the dependence of the influence of the leader on the follower dynamics. This dependence is shown in Fig.~\eqref{fig:ATsnaps}. The plots are drawn from Eq.~\eqref{eq:phsLockCond} with parameters $r_2=1.0$ \& $\omega_2=3.0$. The dynamics of the leader are assumed to be in steady state and, hence, values of $r_1$ are taken as $\sqrt{\lambda_1}$. The figure shows coloured regions satisfying Eq.~\eqref{eq:phsLockCond} representing phase locking for $1:1$, $2:3$, $1:2$, and $1:3$ synchronized dynamics. The \textit{tongues} can be connected to the x-axis, as with $1:1$ case in all the plots or with $1:2$ synchronization in Fig.~\eqref{fig:ATsnaps}(iv, v, vi) and with $2:3$ synchronization in Fig.~\eqref{fig:ATsnaps}(xi, xii). This is the saddle-node bifurcation point denoting the origin of a stable limit cycle. The \textit{tongues} can be disconnected, \textit{floating tongues}, common with mostly higher-order synchronizations. The disconnected \textit{tongues} show the requirement of higher leader amplitude or stronger coupling for locking. These are essentially because of the non-identical nature of the leader and follower. Along the edge of the \textit{tongues} if the boundary separates phase-locked and unlocked regions, the systems undergo Neimark-Sacker bifurcation resulting in a quasi-periodic behavior. This is not possible in the overlap regions, usually with the \textit{tongues} of higher order.

In the figure, from left to right, plots are shown with increasing coupling coefficients, and from top the bottom, with increasing $\alpha$. The basic trend among the plots suggests a broader $1:1$ \textit{tongues} and narrower for other $p_1:p_2$ ratios. Further, the coupling stretches the \textit{tongues}, resulting in broader regions in the plane. This suggests that locking regions spread with coupling. This stretching is proportional to the $p_1:p_2$ ratio. The non-isochronicity bends the locking regions for higher $r_1$ values towards higher $\omega_1$ values. The bending follows the values along the resonance conditions, to be discussed in the following sections.

\subsection{Amplitude response}\label{SSS-AmpRes}

\begin{figure}[!ht]
\centering
\includegraphics[width=0.8\linewidth]{"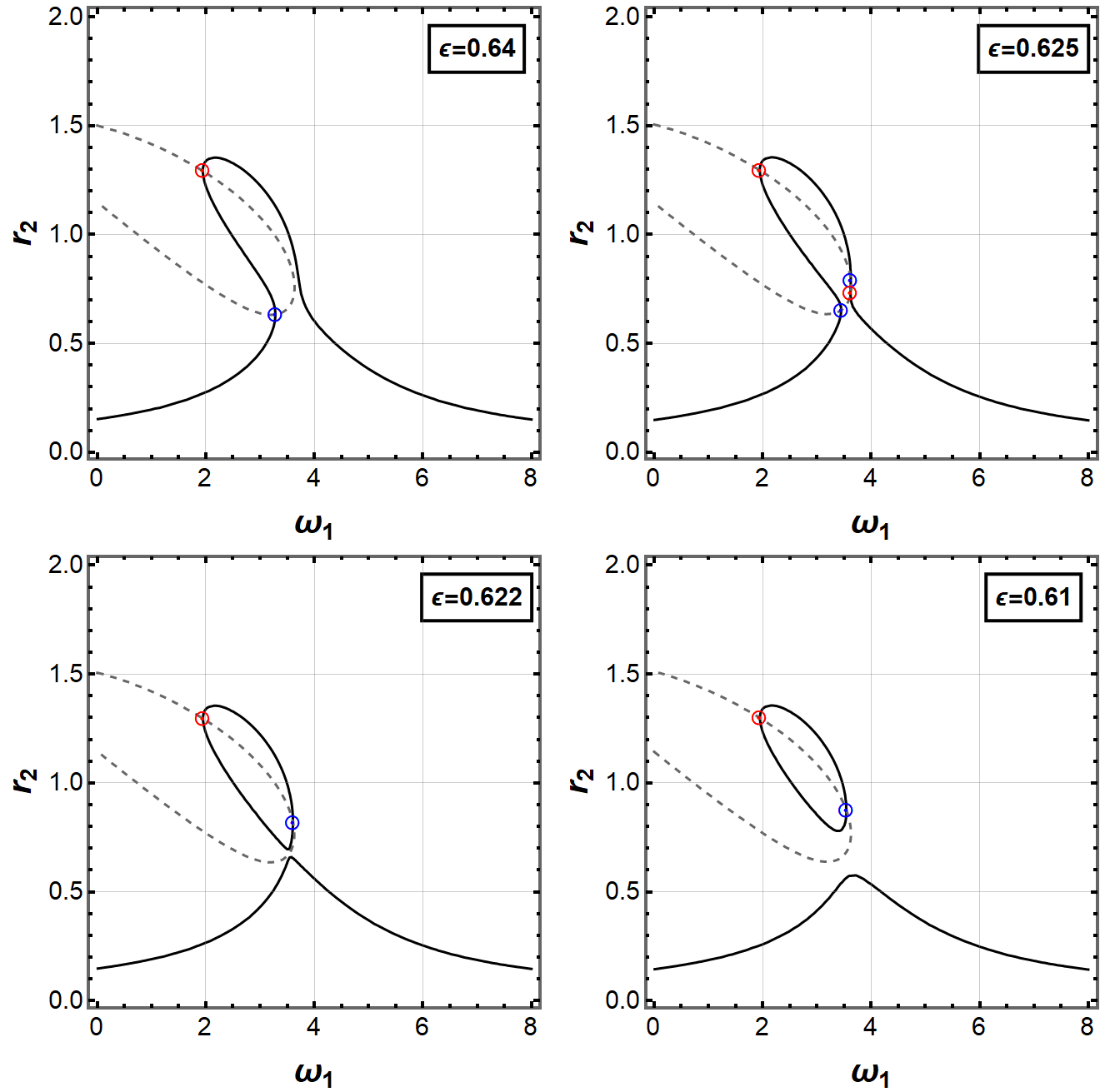"}

\caption{Follower amplitude response to leader frequency. The solid curve denotes the response. The dashed curve represents the fold conditions. Red $\odot$ denotes supercritical saddle-node bifurcation and blue $\odot$ represents subcritical saddle-node bifurcation. Parameters: $\alpha=1.0$, $\omega_2=3.0$, $\lambda_1=1.0$, $\lambda_2=2.0$.  }

\label{fig:freqResp}
\end{figure}

The expression connecting the amplitude with the leader's parameters is given by Eq.~\eqref{eq:amplitudeResp}. Considering a long time limit, such that the leader attains the steady state and oscillates in the limit cycle. For this setting, the dynamics of the follower appear as a forced oscillator. The response of $r_2$ to the variation of leader forcing frequency is shown in Fig.~\eqref{fig:freqResp}. The parametric conditions for the plot are: $\alpha=1.0$, $\omega_2=3.0$, $\lambda_1=1.0$, $\lambda_2=2.0$. The figure demonstrates resonance bebehaviorith respect. $\omega_1$ by showing a peak at the resonance frequency. For positive finite non-isochronicity, $\alpha$, the response curve bends towards lower frequency values, because of which $r_2$ shows multiple fixed points for the same frequency. 
The points around which the curve bends are called \textit{fold} points, formed as the system undergoes \textit{fold} bifurcation or saddle-node bifurcation. The conditions for the bifurcation are, from Eq.~\eqref{eq:amplitudeResp},
\bea g =\left((p_1\omega_1 - p_2\omega_2)r_2 - \alpha (p_1 r_1^2 - p_2 r_2^2)r_2 \right)^2 + \left((\lambda_2 - \epsilon  - r_2^2)r_2 \right)^2 - \epsilon^2 r_1^2 = 0, \quad
\frac{\partial g}{\partial r_2} = 0. 
\label{eq:foldCond}
\eea

These bifurcations describe the disappearance or appearance of fixed points with the variation of the parameter. In Fig.~\eqref{fig:freqResp}, the conditions Eq.~\eqref{eq:foldCond} are shown with dashed curves, which intersect the response curve at bifurcation points. Along increasing frequency, the bifurcation points after which a stable and an unstable branch originate are called supercritical saddle-node bifurcations and are denoted by red unfilled markers. For bifurcations where a stable branch and an unstable branch annihilate each other are called subcritical saddle-node bifurcations and are denoted by blue markers. The multistability region between the two bifurcations gives an impression of the presence of hysteresis in the system; however, as will be explained in the next section, the region seldom shows hysteresis. \\
In the figure, the plots are arranged with decreasing coupling strengths. At a critical value of coupling, the system undergoes \textit{isola} bifurcation, resulting in an isolated response curve, referred to as \textit{isola}. The conditions for the \textit{isola} bifurcation, in addition to Eq.~\eqref{eq:foldCond}, are
\be \frac{\partial g}{\partial \omega_1}=0, \,\,\,\,
\frac{\partial^2 g}{\partial r_2^2} \neq 0.
\label{eq:isolarwCond}
\ee 
Using the above conditions, the \textit{isola} bifurcation occurs at $r_2=0.67765$, $\omega_1=3.54079$, at a coupling strength $\epsilon=0.62237$. 

\begin{figure}[!ht]
\centering
\includegraphics[width=0.8\linewidth]{"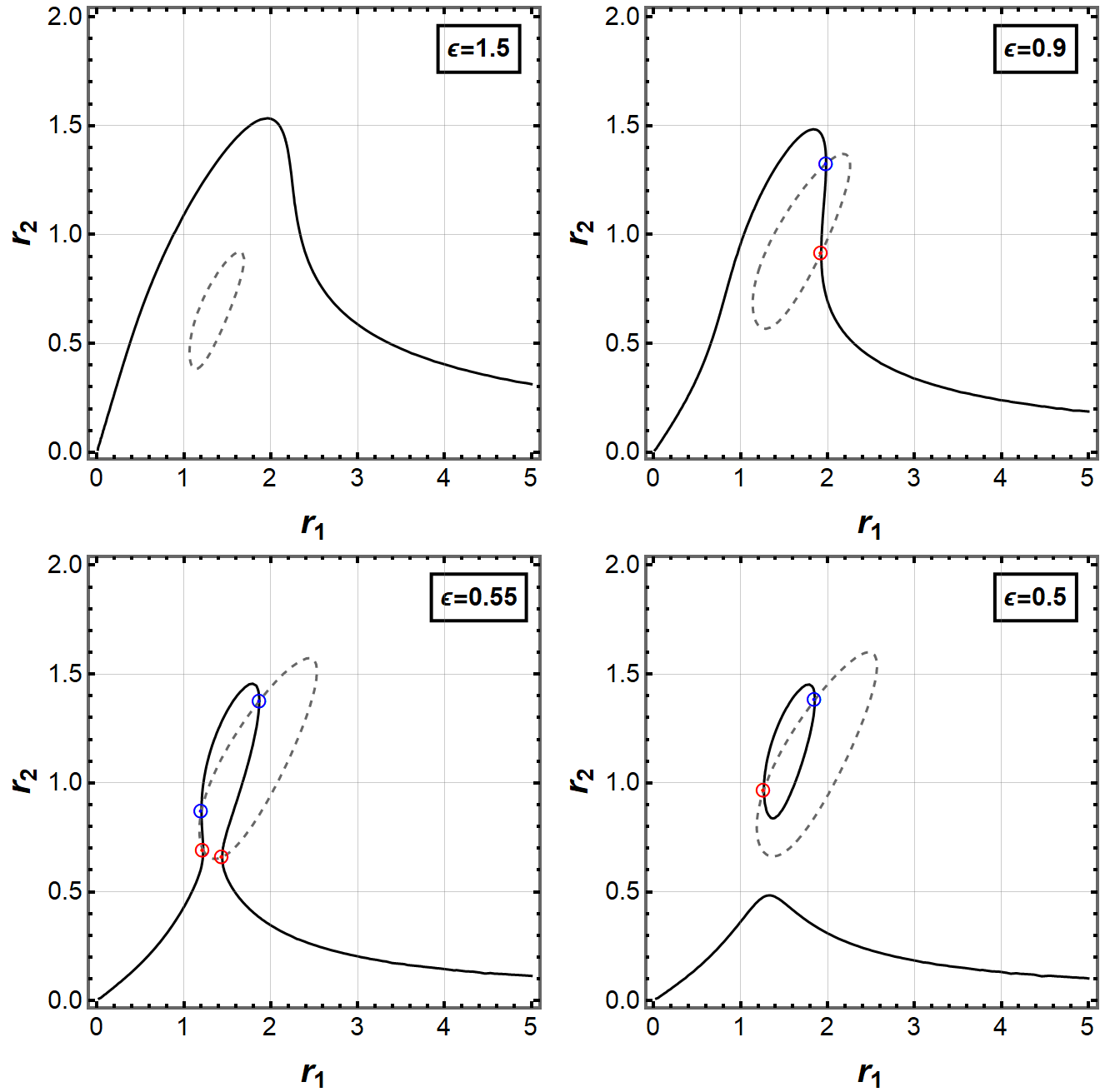"}

\caption{Follower amplitude response to leader amplitude. The solid curve denotes the response. The dashed curve represents the fold conditions. Red $\odot$ denotes supercritical saddle-node bifurcation and blue $\odot$ represents subcritical saddle-node bifurcation. Parameters: $\alpha=1.0$, $\omega_1=3.0$, $\omega_2=2.0$, $\lambda_2=2.0$. }

\label{fig:ampResp}
\end{figure}

The response of the follower amplitude to the leader's amplitude also shows resonance behavior with a peak at the resonant $r_1$ value. The reason for resonance in the 
The response is shown in Fig.~\eqref{fig:ampResp} for parameter values: $\alpha=1.0$, $\omega_1=3.0$, $\omega_2=2.0$, $\lambda_2=2.0$. The figure shows plots with decreasing coupling strength leading to the formation of \textit{isola}. The conditions for the saddle-node bifurcations are given by Eq.~\eqref{eq:foldCond} as before; however, this time to be solved for $r_1$ and $r_2$. The extra conditions for \textit{isola bifurcations} are given by
\be \frac{\partial g}{\partial r_1}=0, \,\,\,\,
\frac{\partial^2 g}{\partial r_2^2} \neq 0.
\label{eq:isolarrCond}
\ee  
The \textit{isola} bifurcation occurs at $r_1=1.32842$, $r_2=0.65649$ at the critical coupling strength of $\epsilon=0.53633$. Further decreasing the coupling strength results in the vanishing of the \textit{isola}.

\subsubsection{Mechanism of resonance and isola}\label{SSS-MechResoIsola}

The resonance in the amplitude-frequency response for $1:1$ synchronized state comes due to the maximization of the $cosine$ part at that point. At this point, the energy from leader to follower dynamics is maximized. However, energy is not uniformly transferred at all amplitudes, $r_2$. As the coupling coefficient is decreased, the difference in the energy transfer becomes more apparent. Finally, at the critical point, the response is divided into two segments forming an \textit{isola}. For the main branch (lower part), lower coupling will lead to a decrease in the amplitude, and for the \textit{isola} (upper branch), it results in an increment.

The mechanism of resonance in amplitude-amplitude response is different. In this case, the resonance is the result of the combined effects of coupling and non-isochronicity. For finite coupling in a synchronized state, the effective frequency of the follower changes with the amplitude and is amplified by $\alpha$. For $\alpha=0$ the $r_2$ will increase monotonously with $r_1$. However, as $\alpha$ becomes finite, there will be an $r_1$ value for which the effective frequency corresponds to a resonance maximum. This gain in energy is amplitude dependent, as in the previous case, resulting in a segregation of the response curve forming an \textit{isola}.

\section{Qualitative stability properties}\label{StabilityAnalysis}
The analytical results reported in the previous section do not provide the complete dynamical picture. To understand the follower's dynamics, the stability of its steady states needs to be calculated. In this section, we will analyze qualitatively the long-time behavior of the system through the stability of its fixed points.

To proceed, consider the Jacobian of the follower's amplitude and phase difference dynamics at the point $(r_2^*,\,\, \phi^*)$,
\bea
J = 
\begin{pmatrix}
\lambda_2 - \epsilon - 3 {r_2^*}^2 & \epsilon r_1 \sin(\phi^*) \\
-2 \alpha r_2^* - \epsilon \frac{r_1}{r_2^*} \sin(\phi^*) & -\epsilon \frac{r_1}{r_2^*} \cos(\phi^*)
\label{eq:JacobMat}
\end{pmatrix}.
\eea

\begin{figure}[!ht]
\centering
\includegraphics[width=0.8\linewidth]{"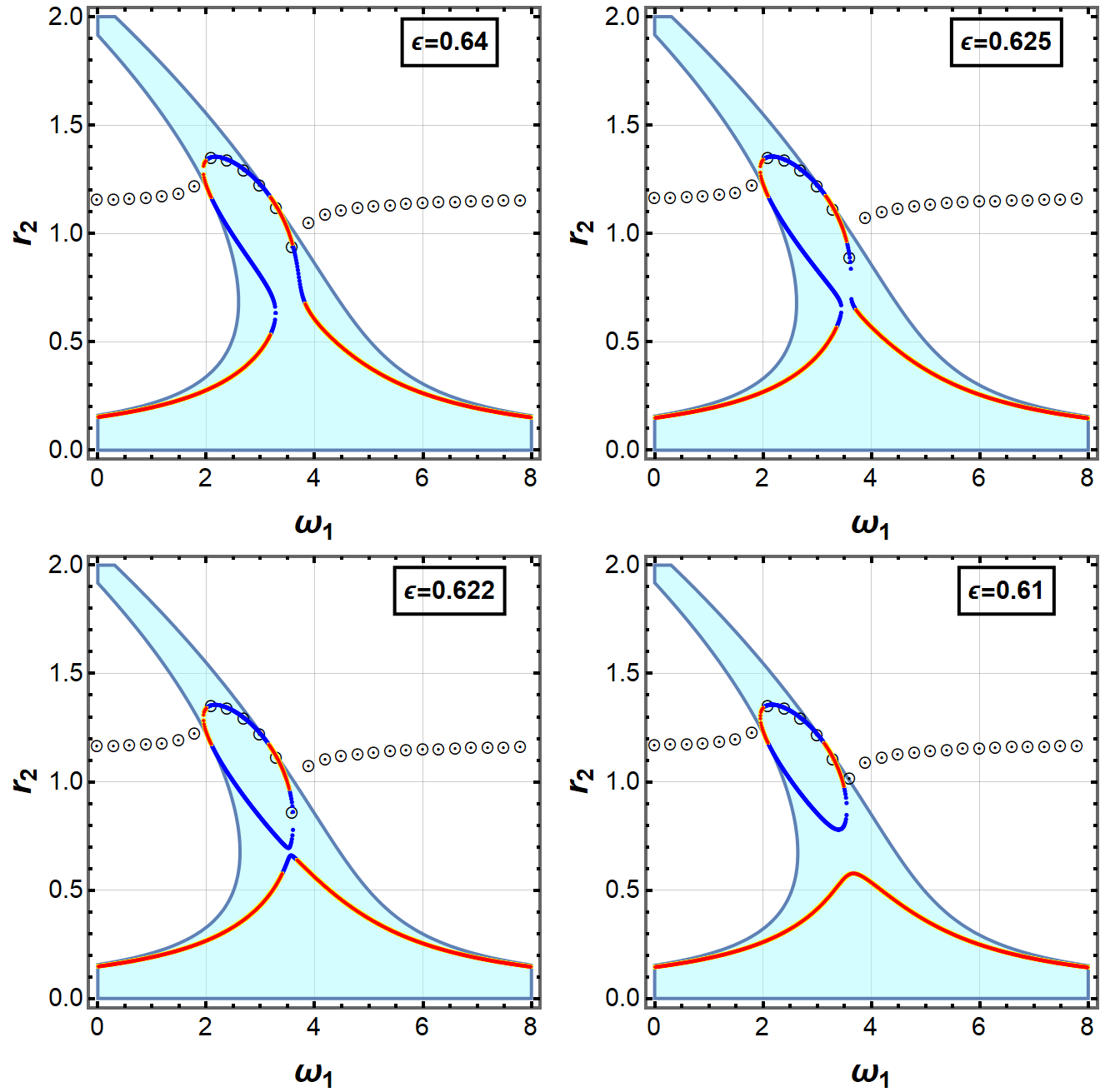"}

\caption{Illustration of stability in the amplitude response to the leader's frequency. The blue shaded regions represent phase locking. Red data points represent unstable (or saddle) responses, and blue data points represent stable responses. Markers $\odot$ represent numerically obtained values. Parameters: $\alpha=1.0$, $r_1=1.0$, $\omega_2=3.0$, $\lambda_2=2.0$.}

\label{fig:freqRespStab}
\end{figure}

\begin{figure}[!t]
\centering
\includegraphics[width=0.8\textwidth]{"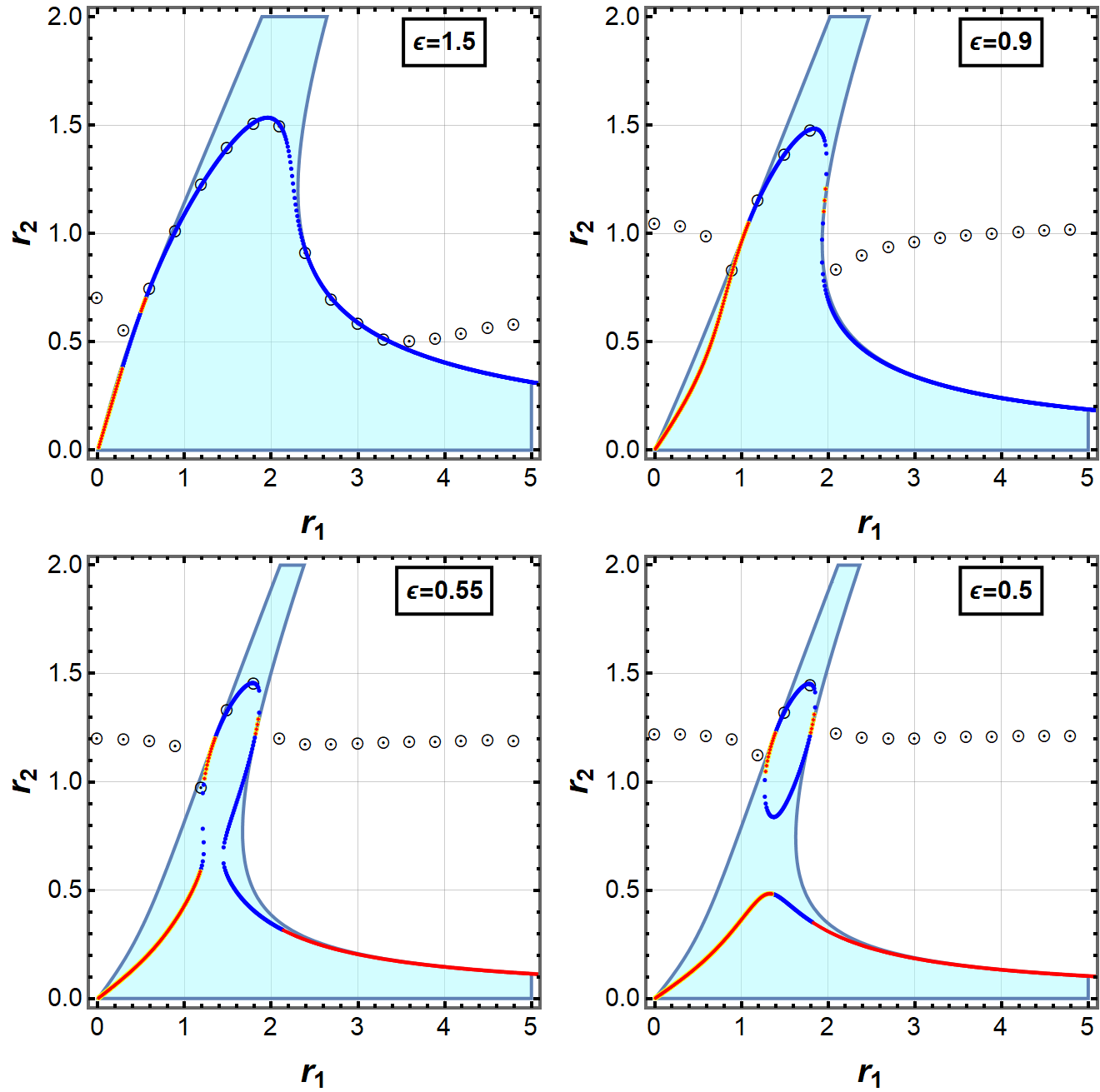"}

\caption{Illustration of stability in the amplitude response to the leader's amplitude. The blue shaded regions represent phase locking. Red data points represent unstable (or saddle) responses, and blue data points represent stable responses. Markers $\odot$ represent numerically obtained values. Parameters: $\alpha=1.0$, $\omega_1=3.0$, $\omega_2=2.0$, $\lambda_2=2.0$. }

\label{fig:ampRespStab}
\end{figure}

The eigenvalues of this matrix determine the stability of the fixed points. The fixed points of the amplitude dynamics are given in the previous section as the amplitude response. The projection of stability mapping over the response is given in Fig.~\eqref{fig:freqRespStab} and Fig.~\eqref{fig:ampRespStab}. 
The figures are plotted with the same parameter settings as in the response plot. In the figures, the blue shaded regions are $1:1$ phase-locked. It is obvious that the fixed points (responses) reside inside the locked regions. However, the phase-locked regions are not entirely stable. The stable and unstable parts of the responses are shown in the figures with blue and red colors, respectively. The loss of stability of $r_2$ can be understood from the Jacobian Eq.~\eqref{eq:JacobMat}. The first element of the matrix gives the stability of the amplitude with the variation of $r_2$. Clearly, $r_2^*$ is unstable for lower values, particularly, $\lambda_2-\epsilon>3{r_2^*}^2$. This is also evident from the plots, especially for weak coupling. Instability from these points often results in stability of higher $r_2$ values within the phase-locked states. But this is not the general case. Amplitude fixed points can also lose stability because of their dependence on the phase. Small perturbations to the phase at the boundary of the \textit{Arnold's tongue} perturb the amplitude response, as well. This change feeds back to the phase dynamics through the non-isochronous term, which stabilizes the phase-drift states further. This feedback loop mechanism creates an oscillating amplitude dynamics away from the response. 

The numerical results are obtained using long-time averaging of the amplitudes by solving Eq.~\eqref{eq:cSLOpolarr2} and Eq.~\eqref{eq:cSLOpolartheta2}. These points are shown with $\odot$ markers in the plot. These points correspond to quasiperiodic oscillations of the system with incommensurate frequencies. These solutions will be discussed in the next section.

Further observations from numerical experiments reveal that stability is concentrated near the resonance region, which shrinks with decreasing coupling coefficient. For weak coupling, for responses, at and near resonance, the dissipation in the system balances forcing, which is mainly caused by coupling. These points are well inside the tongues, keeping them stable under perturbations. Away from the resonance, the system is near the boundary of the tongue, where the dissipation in the system interacts with the nonlinearity, causing small perturbations to sustain and rise towards quasiperiodic oscillations. Further, stronger coupling leads to more stable phase-locking as inferred from Fig.~\eqref{fig:ampRespStab}. For lower coupling strength, the higher leader amplitude corresponds to unstable phase-locked states, depicted for $\epsilon=0.5$ in the figure.

\section{Quasiperiodic oscillations}\label{QuasiperiodOsci}

The nonlinear interactions, together with coupling, create higher-order modes in the system. This is discussed previously in the phase response section. As the linear mode loses stability, the system leaves the periodic attractor and starts evolving on a torus. The torus is typically a 2-torus corresponding to two frequencies. As seen through numerical experiments, higher-order modes are unstable in the follower's dynamics. Therefore, the torus is formed by incommensurate frequencies, and the solutions that evolve on the torus are quasiperiodic oscillations.

\begin{figure}[!h]
\centering
\includegraphics[width=1\linewidth]{"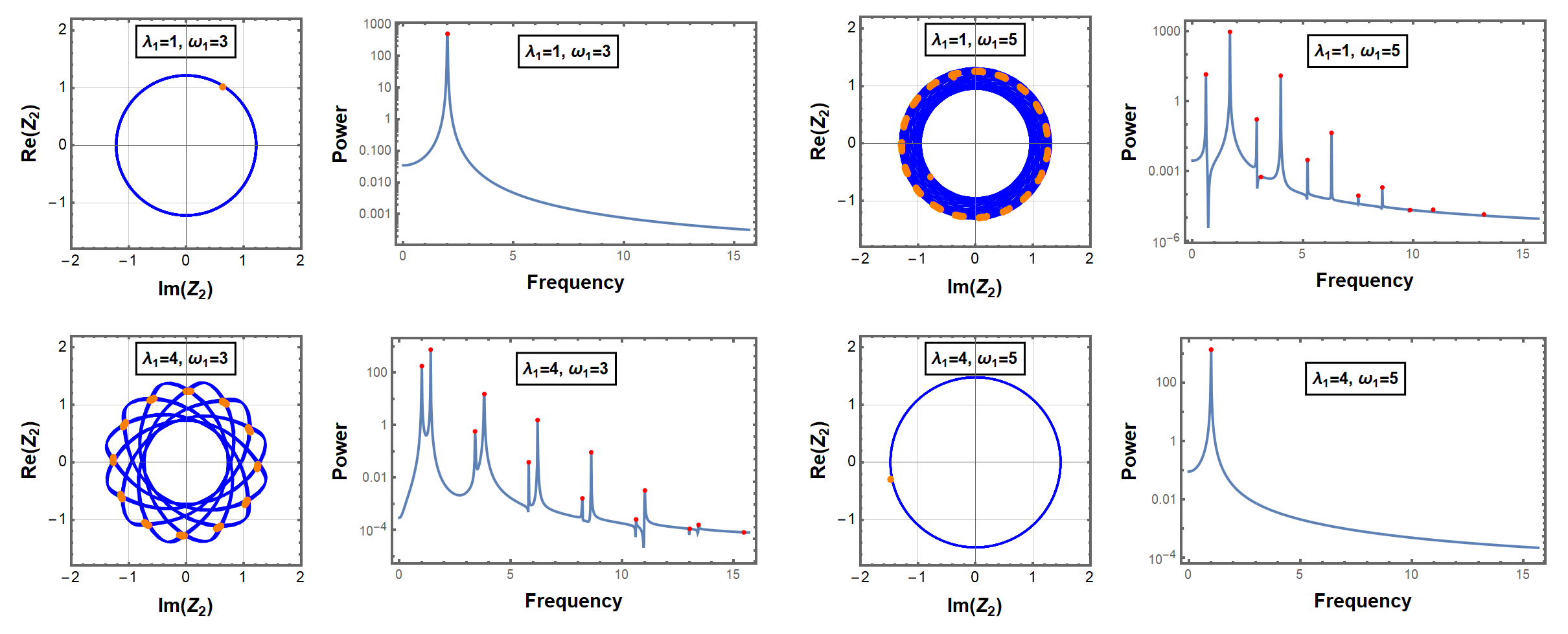"}

\caption{Illustration of quasiperiodic oscillations in the follower's dynamics. The plot depicts four scenarios with a phase portrait and a power spectrum. The variable parameters are shown in the insets. Orange data points in the phase portrait represent Poincar\`e section, and red dots in the power spectrum are peaks. Parameters: $\alpha=1.0$, $\omega_1=3.0$, $\omega_2=2.0$, $\lambda_2=2.0$. }

\label{fig:QuasiPerio}
\end{figure}

The quasiperiodic solutions reside in the phase drift regions outside the tongues. As discussed earlier, with the system's departure from near-resonance values, phase-locked states lose stability to higher-order modes. However, these modes are also unstable, leading to irrational frequency combinations to take place. These solutions densely fill the torus as they evolve with time. The formed torus remains invariant in the presence of small enough perturbations, denoting its stability. This behavior is depicted in Fig.~\eqref{fig:QuasiPerio} with four scenarios. The non-varying parametric conditions are chosen to match previous analysis:  $\alpha=1.0$, $\omega_2=2.0$, $\lambda_2=2.0$. The scenarios are shown with a phase portrait and power spectrum. The power spectrum is obtained using the \texttt{Periodogram} function in \textsl{Mathematica} software. The function, essentially, gives the squared magnitude of the discrete Fourier transform of the signal. The peaks are identified and labeled with red dots. Further, the phase portraits are accompanied by orange dot points representing the Poincaré sections obtained from the numerical simulations.

In the first scenario, plots are obtained for $\lambda_1=1.0$ and $\omega_1=3.0$, which lie in the stable phase-locked zone. The plot shows a limit cycle with a unique peak in the power spectrum. With increasing $\omega_1$, the system moves to the phase-drift region. This is shown in the second scenario with $\lambda_1=1.0$ and $\omega_1=5.0$. Here, the quasiperiodic behavior is depicted with multiple incommensurate frequency peaks. In the bottom two scenarios, the phase-drift and phase-locked scenarios are depicted for $\omega_1=3.0$ and $\omega_1=5.0$ for $\lambda_1 = 4.0$.

\begin{figure}[!h]
\centering
\includegraphics[width=1\textwidth, keepaspectratio]{"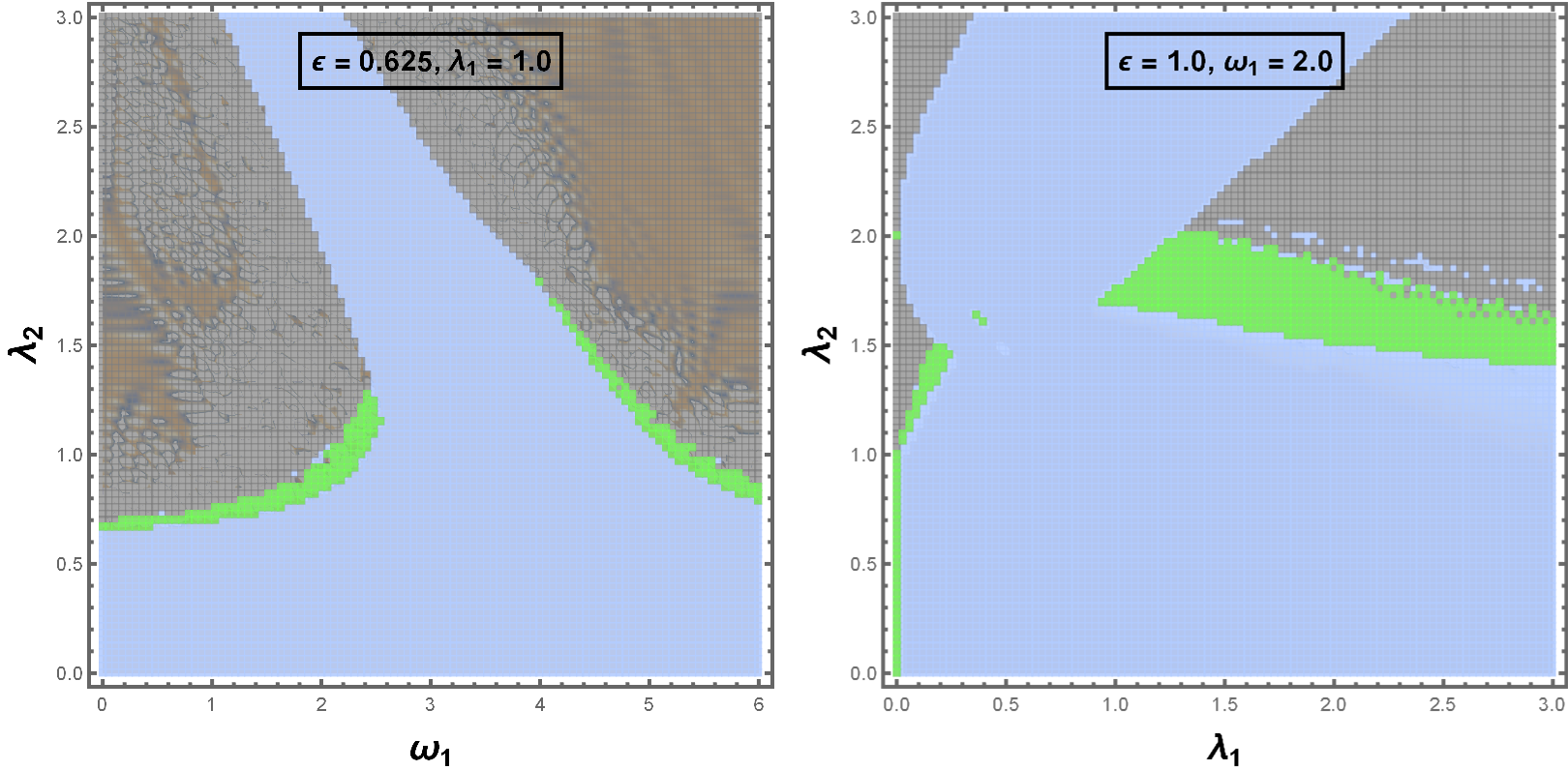"}

\caption{Maps of Lyapunov exponents illustrating periodic and quasiperiodic attractors in the follower's dynamics and its dependence on the leader's parameters. The plot shows parameter regions for phase locked (blue), phase drift (gray), and partial quasiperiodic synchronization (green) in $\lambda_2 - \omega_1$ (left) and $\lambda_2 - \lambda_1$ (right) parameter plane. The coupling strength is shown in the inset. Parameters: $\alpha=1.0$, $\omega_2=3.0$. }

\label{fig:QPvsNQPFl2Lwl}
\end{figure}

\subsection{Conditional dependence of quasiperiodicity}\label{QO-paraDepend}

The stability of periodic and quasiperiodic attractors depends on multiple factors. In this section, we will explore the stability of these attractors over parameter space.

The effects of the leader's parameters on the follower's attractors are given in Fig.~\eqref{fig:QPvsNQPFl2Lwl}. The plots are \textit{maps of Lyapunov exponents} obtained by calculating Lyapunov exponents over the grid of values in the parameter space. These charts are superimposed over the exponent densities, giving the texture representing the strength of the exponents. 
 The figure shows two plots in parameter space of $\omega_1-\lambda_2$ and $\lambda_1-\lambda_2$. For periodic oscillations, perturbations do not grow or decay along the trajectories; hence, the largest exponent is zero. These are denoted by blue in the plots. In the quasiperiodic case, there are more than one zero exponent; however, there is sub-exponential growth along the trajectory. Therefore, due to finite-time approximation in the calculations, the exponents are near zero, $|LE|<10^{-6}$, rather than zero. These are shown in gray in the plots. Further, the plots are filtered with long-time phase difference information for each parameter value. Parameter values near the bifurcation boundaries show persistence of synchronization along one of the axes of the torus. For these regions, the phase difference shows a converging trend and remains locked. This show \textit{partial quasiperiodic synchronization} represented by green in the plots.    
 The parametric conditions are kept the same as the analysis done so far. The first plot shows stable quasiperiodic behavior except for low $\lambda_2$ values and along the resonance. This is in conformity with the previous results. The robustness of stable periodic oscillations along the resonance can be attributed to the fact that, along resonance, the amplitude dynamics remain weakly affected by the phase perturbations. Hence, the amplitude settles to the fixed point defined largely by $\lambda_2$, and the feedback to the phase dynamics through $\alpha$ also stabilizes the phase dynamics. This behavior is also visible in the second plot, where periodic dynamics remain stable along resonance in the $\lambda_1-\lambda_2$ plane. The resonance with respect to $\lambda_1$ essentially arises through coupling and amplitude dependence in phase dynamics. Around the $\lambda_2$ values for which resonance occurs, perturbations below certain magnitudes are suppressed. This magnitude is determined by the coupling strength. This is evident from the second plot, where, periodic window narrows with decreasing $\epsilon$. The smaller values of $\lambda_2$ show robustness because weaker amplitudes are not enough to sustain the perturbations in the feedback mechanism.

\begin{figure}[!h]
\centering
\includegraphics[width=1\textwidth, keepaspectratio]{"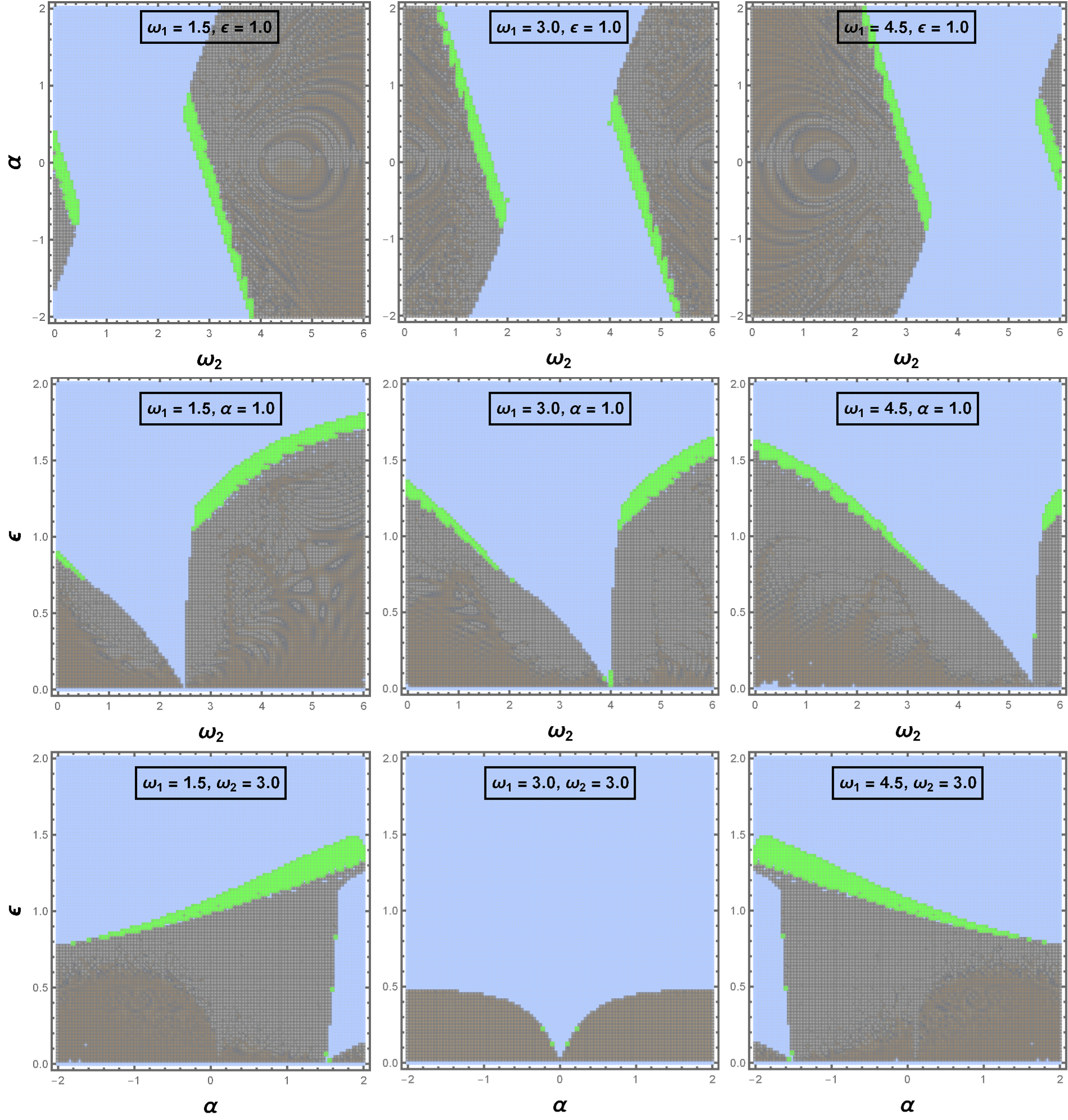"}

\caption{Maps of Lyapunov exponents over followers' parameter space. The plot classifies parameter regions for which the dynamics show phase locked (blue), phase drift (gray), and partial quasiperiodic synchronization (green) in 
$\alpha-\omega_2$ plane (first row), $\epsilon-\omega_2$ plane (second row), and $\epsilon-\alpha$ plane (third row). Each row contains plots for $\omega_1 = \{1.5, 3.0, 4.5\}$ . The varying parameter values are given in the insets. Fixed Parameters: $\lambda_1=1.0$, $\lambda_2=2.0$. }

\label{fig:QPvsNQPFeaw}
\end{figure}

The response of the follower to the leader's forcing can also be attributed to its own parameter values. In previous discussions, we have emphasized the role of coupling, non-isochronicity, and frequency in determining the steady states and their stability. A general dependence over the expanse of these parameters is given in Fig.~\eqref{fig:QPvsNQPFeaw}. The figure shows plots of charts of Lyapunov exponent over combinations of $\alpha$, $\epsilon$, and $\omega_2$. For each combination, the plots are drawn corresponding to three leader frequencies, $\omega_1\in \{1.5,3.0,4.5\}$. The fixed parameters are: $\lambda_1=1.0$, $\lambda_2=2.0$. The plots show two regions - quasiperiodic (gray) and periodic (blue).

For the analysis, it is convenient to consider the relative frequency $\omega_{21}=\omega_2/\omega_1$. For fixed coupling, the distribution of periodic and quasiperiodic solutions over $\alpha-\omega_2$ parameters is shown in the first row. The plots show a periodic region around $\omega_{21}=1$ which shifts towards higher $\omega_2$ with increasing $\omega_1$. The dependence of periodic stability on $\alpha$ is asymmetric around $\omega_{21}=1$. The negative values of $\alpha$ increase the effective frequency of the follower; therefore, the periodic synchronization state is achieved for lower values of $\omega_1$. However, as $\alpha$ is increased, periodic states require higher $\omega_1$ values, which, ultimately, reach a maximum value and start decreasing afterward. For $\omega_{21}>1$, the difference between the leader and follower frequency is negative; hence, for more negative $\alpha$ values, periodic behaviour shows resilience to the perturbations for higher frequencies. As the $\alpha$ values are increased in this region, the resilient frequency decreases till a minimum and then increases again.

These bounds are regulated by the coupling strength; a lower $\epsilon$ corresponds to a narrower periodic region with extrema coming towards $\alpha=0$. Higher values result in broader blue regions with extrema going farther from the center. This can be deduced from the other two rows of Fig.~\eqref{fig:QPvsNQPFeaw}, where higher $\epsilon$ corresponds to greater stability of periodic solutions. The strong coupling, essentially, helps in the robust transfer of energy between oscillators, keeping it stable. The second row displays the \textit{Arnold's tongue} in $\omega_2-\epsilon$ plane. Outside the tongues are parameters for quasiperiodic oscillations. The third row illustrates symmetric parametric conditions across the $1:1$ synchronized state in $\alpha-\epsilon$ parameter space. Further, for $\omega_{21}<1$ higher non-isochronicity demands greater coupling strength for stable periodic oscillations, while, for $\omega_{21}>1$ it is reversed.

\section{Triadic Stuart–Landau network}\label{TriadSLO}

So far, we have considered two oscillators with unidirectional coupling. This paradigm, essentially, gave conditions of a forced Stuart-Landau. This forcing is periodic in nature. However, in many situations, such as in a chain of oscillators, the effective forcing could be quasiperiodic. To implement this, we shall add another oscillator, $Z_3$, to the existing system. In this section, we will study three coupled Stuart-Landau oscillators with the objective of understanding $Z_3$'s response to quasiperiodic forcing from $Z_2$ oscillations.

The new system can be summarised as 

\bea
\ddot{Z_1} = (\lambda_1 + \iota \omega_1 -(1 + \iota \alpha)|Z_1|^2)Z_1,  \\
\ddot{Z_2} = (\lambda_2 + \iota \omega_2 - (1 + \iota \alpha)|Z_2|^2)Z_2 +   \epsilon(Z_1 - Z_2), \\
\ddot{Z_3} = (\lambda_3 + \iota \omega_3 - (1 + \iota \alpha)|Z_3|^2)Z_3 +   \epsilon(Z_2 - Z_3). 
\label{eq:3cSLOmain}
\eea

The corresponding equations in polar form are given as

\bea
\dot{r_1} = (\lambda_1 - r_1^2)r_1, 
\label{eq:3cSLOpolarr1} \\
\dot{r_2} = (\lambda_2 - \epsilon  - r_2^2)r_2 + 
\epsilon r_1 \cos(\theta_1 - \theta_2),
\label{eq:3cSLOpolarr2}  \\
\dot{r_3} = (\lambda_3 - \epsilon  - r_3^2)r_3 + 
\epsilon r_2 \cos(\theta_2 - \theta_3),
\label{eq:3cSLOpolarr2}  \\
\dot{\theta_1} = \omega_1 - \alpha r_1^2,
\label{eq:3cSLOpolartheta1}  \\
\dot{\theta_2} = \omega_2 - \alpha r_2^2 + \epsilon\frac{r_1}{r_2}\sin(\theta_1 - \theta_2), 
\label{eq:3cSLOpolartheta2} \\
\dot{\theta_3} = \omega_3 - \alpha r_3^2 + \epsilon\frac{r_2}{r_3}\sin(\theta_2 - \theta_3). 
\label{eq:3cSLOpolartheta2}
\eea

As the oscillator $Z_3$ is unidirectionally coupled to $Z_2$, the analytical deductions shall be the same as in Eq.~\eqref{eq:amplitudeResp} and Eq.~\eqref{eq:phaseResp}. 

\subsection{Response landscapes under quasiperiodic driving}\label{TSLO-Resp}

In periodic forcing, the oscillator locks its phase to the forcing signal. This locking could be $1:1$ with forcing or with other ratios. In the cases considered so far, higher-order locked states were unstable and could be achieved in the dynamics. In the case of quasiperiodic forcing, the forcing is composed of multiple incommensurate phases. This makes the locking difficult for the forced oscillator. The oscillator may lock to one of the phases and get modulated by the other. The oscillator, eventually, evolves to a torus attractor to accommodate multiple phases. This renders the system to show quasiperiodic oscillation. However, with increasing coupling strength or forcing, together with nonlinearities, the torus becomes unstable and breaks down, giving rise to scores of complex dynamical phenomena.

\begin{figure}[!th]
\centering
\includegraphics[width=1\textwidth]{"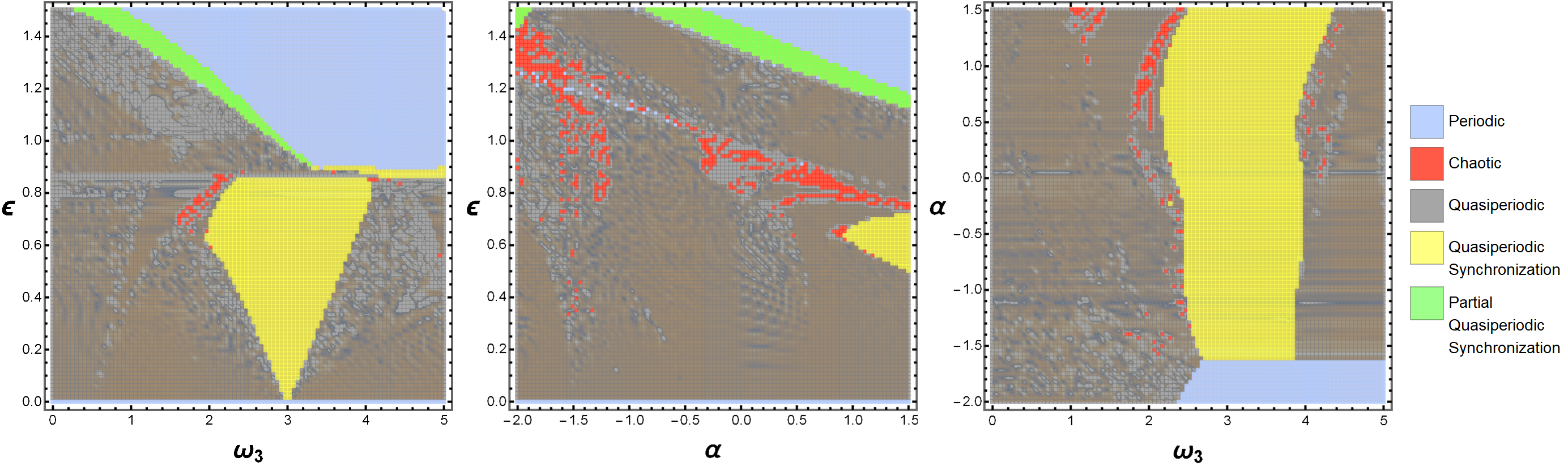"}

\caption{Illustration of chaos in three parameter space: $\omega_3-\epsilon$, $\alpha-\epsilon$, $\omega_3-\alpha$. The plots consist of periodic synchronizations (blue), quasiperiodic non-synchronized values (gray), quasiperiodic synchronized values (yellow), partial quasiperiodic synchronization (green), and chaotic values (red). Variable parameters - First plot: $\alpha=1.0$, second plot: $\omega_3=2.0$, third plot: $\epsilon=0.8$. Fixed Parameters: $\lambda_1=1.0$, $\lambda_2=2.0$, $\lambda_3=2.0$, $\omega_1=4.5$, $\omega_2=3.0$. }

\label{fig:ChaosParam}
\end{figure}

In this work, we will stick to the parameter ranges discussed so far. The summarised results of this section are given in Fig.~\eqref{fig:ChaosParam}. The figure consists of three plots in the parametric space of $\omega_3-\epsilon$, $\alpha-\epsilon$, and $\omega_3-\alpha$. The plots are \textit{maps of Lyapunov exponents} obtained by calculating Lyapunov exponents over the grid of values in the parameter space. These charts are superimposed over the exponent densities, giving the texture representing the strength of the exponents. For some values, the exponents show, erroneously, high exponents, particularly at boundaries of bifurcations. Therefore, the charts are filtered by the spectrum entropy obtained using densities of the power spectrum at those points. The fixed parameters are: $\lambda_1=1.0$, $\lambda_2=2.0$, $\lambda_3=2.0$, $\omega_1=4.5$, $\omega_2=3.0$. These values are selected by taking into account Fig.~\eqref{fig:QPvsNQPFeaw} so that the influence of $Z_2$ ranges from periodic to quasiperiodic. 

The plots show the presence of multiple new attractors. When the parameters in the plots belong to a periodic oscillation of $Z_2$, the response of $Z_3$ is to either synchronize periodically with the forcing or evolve over a torus displaying quasiperiodic oscillation. The periodically synchronized oscillation part is shown in the plots in blue. These points are mainly found in regions of strong coupling and high frequency. The quasiperiodic responses are coded in gray and yellow. For parameters corresponding to quasiperiodic oscillations of $Z_2$, the interactions result in quasiperiodic and chaotic responses. 

\subsection{Quasiperiodic synchronisation}\label{TSLO-qausiSync}

With quasiperiodic influence, the oscillator is forced to evolve on a torus. The axes of the torus are incommensurate frequencies of the quasiperiodic oscillations. These frequencies can lock onto each forcing frequency separately or to a combination of them. This synchronizes the dynamics of the two oscillators, and the corresponding phase differences oscillate in a bounded domain. These oscillations are termed as \textit{quasiperiodic synchronizations} (QS) and are represented with yellow markers in the plots of the Fig.~\eqref{fig:ChaosParam}. The locked regions in the parameter space form \textit{Arnold's tongues}. In the plot, higher-order tongues are not shown.

Sometimes, in the quasiperiodic response to periodic forcing, one of the frequency locks to the periodic forcing. The dynamics then show partial synchronization along one of the axes of the torus and modulate along others. This is referred to as \textit{partial quasiperiodic synchronization}. These regions are found at the boundaries of periodic (blue) and quasiperiodic (gray), and are represented in green in the plots. 

There are some parameter values for which $Z_3$ shows periodic responses to quasiperiodic forcing from $Z_2$. These are found in higher-order synchronized oscillations. Such oscillations are formed when the rational combination of quasiperiodic forcing frequencies is an integer multiple of the forced frequency. Due to limitations of scope, we shall not discuss these oscillations in this work. 

\subsection{Emergence of chaos}\label{TSLO-chaos}

\begin{figure}[!h]
\centering
\includegraphics[width=1\textwidth]{"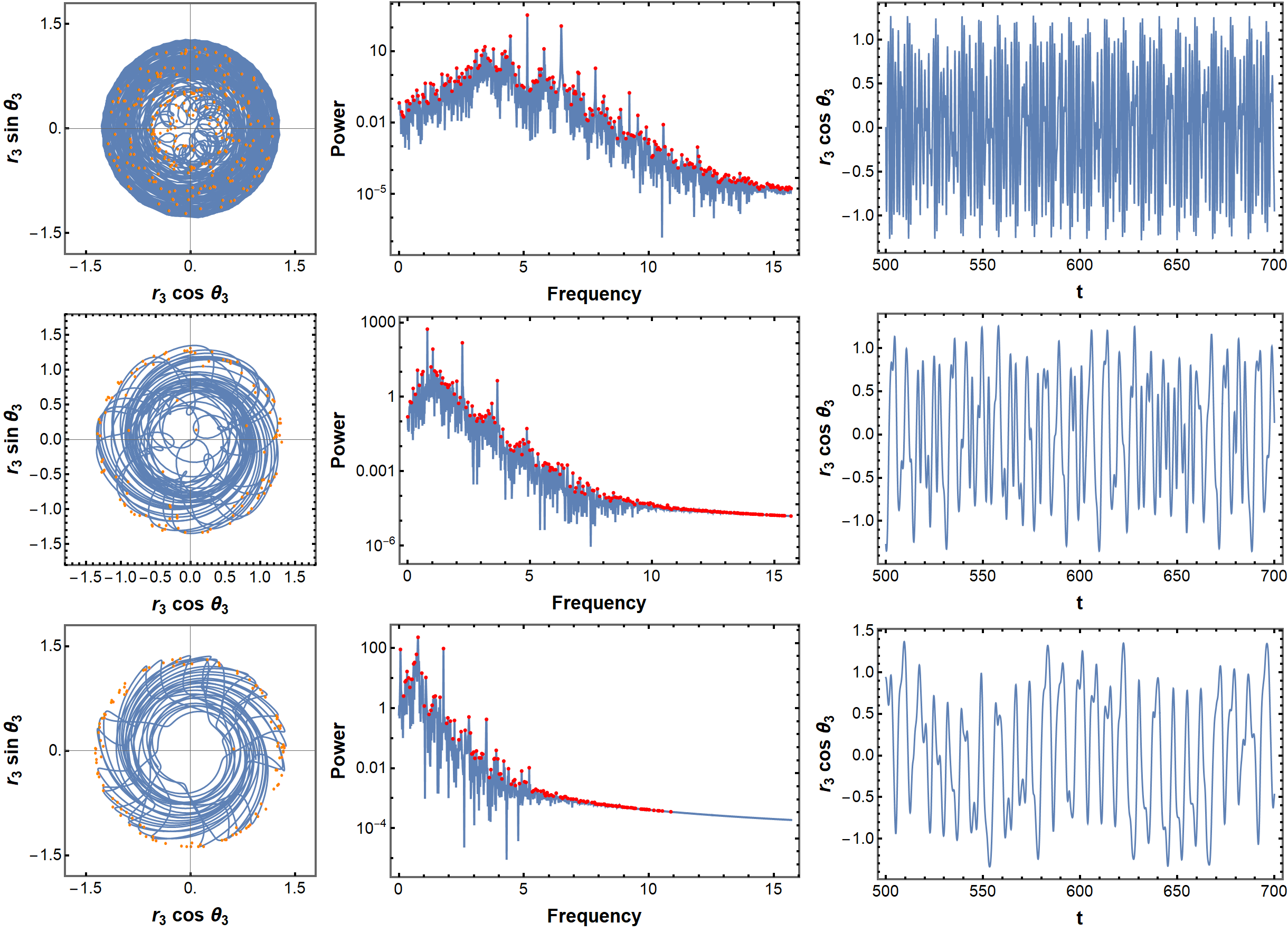"}

\caption{Illustration of chaos over-sampled parametric conditions. The figure consists of three rows corresponding to three parametric conditions for chaotic solutions. The rows depict phase portraits combined with Poincaré sections (orange points), power spectrum with red peaks, and time series.  Parametric conditions - First row: $\omega_3=2.0, \epsilon=1.3, \alpha=-2.0$, second row: $\omega_3=1.86869, \epsilon=0.8, \alpha=0.828283$, third row: $\omega_3=1.76768, \epsilon=0.651515, \alpha=1.0$. Fixed Parameters: $\lambda_1=1.0$, $\lambda_2=2.0$, $\lambda_3=2.0$, $\omega_1=4.5$, $\omega_2=3.0$. }

\label{fig:ChaosSample}
\end{figure}

In the parameter space, the QS locked regions overlap. For parameters in the overlapped regions, the quasiperiodic frequencies compete over the torus attractor. Together with nonlinearity in amplitude-phase dynamics, the torus wrinkles and breaks down, resulting in the formation of a chaotic attractor. Parameter values for chaotic dynamics are displayed in red in the plots. These regions coincide with high coupling coefficients. Depiction of chaotic dynamics is shown in Fig.~\eqref{fig:ChaosSample}. For the plots, three parametric conditions are sampled from each parameter space of Fig.~\eqref{fig:ChaosParam}. For each condition, the plot shows a phase portrait, a power spectrum, and a time series of the dynamics. The phase portrait is superimposed with Poincaré section points. These points are scattered over the phase space, denoting the mixing in the phase space. The power spectrum shows a densely packed, noisy peak. These are common signatures of chaotic dynamics. 

\section{Summary and discussions}\label{Summary}

In this work, we explored the chaotic dynamics in unidirectionally coupled Stuart-Landau oscillators, where the internal dynamics of individual oscillators allow coupling between amplitude and phase through a non-isochronous term. This formalism was referred to as leader-follower dynamics, with the follower responding to the leader's dynamics. The study emphasizes the effects of coupling, non-isochronicity, and frequency on the response dynamics of the follower oscillator. The amplitude death state of the follower shows shifted stability because of the energy pooling in from the leader oscillator. The study of followers' phase dynamics exhibits primary synchronization as well as higher-order synchronization, referred to as \textit{Arnold's tongues}. The shape, size, and position of the tongues are shown to be determined by the interplay of coupling, non-isochronicity, and frequency. However, due to amplitude influence in the phase dynamics, higher-order modes are shown to be unstable. The study of amplitude steady states indicates resonance in both amplitude and frequency responses. These responses undergo \textit{isola} bifurcation, resulting in isolated or detached resonance curves. The isolated branches determine the stable amplitude in the synchronized regions near the resonance. Stability analysis further reveals that periodic attractors corresponding to weak forcing or coupling regimes are dynamically unstable, which pushes the oscillator towards quasiperiodic oscillation on the torus attractor. The mapping of parameter values with the kind of attractor of the oscillator is presented and classified into periodic, quasiperiodic, and partially synchronized quasiperiodic regions. The interplay of the coupling, non-isochronicity, and frequency is analyzed and reported. The parameter studies suggest stability of periodic regions for higher coupling. For weaker coupling, the perturbations persist, giving way to quasiperiodic behavior. The contest between non-isochronicity and frequency determines the resonance conditions and hence the stable periodic regions. 

The addition of another oscillator to the existing framework further increases the complexity in the dynamics. We essentially studied the paradigm where the added oscillator is subjected to quasiperiodic forcing from the leader-follower setting. We found that the added oscillator responds essentially in periodic synchronized oscillation, quasiperiodic oscillation, partially synchronized quasiperiodic oscillations, and chaotic dynamics. 

The results of this study can help in elucidating the formation of dynamical patterns in a chain of oscillators. The qualitative picture portrayed in this work can connect the local dynamics between two oscillators with the global patterns and also shed light on how the selection of parameters leads to transitions in them. However, the study is limited in nature, and many scenarios remain to be answered. The parameter space explored is not exhaustive, and is also limitedly studied in different combinations. Further, the addition of more oscillators could result in more complexity because of the presence of chaos in the system. The amplitude-phase connection can be studied to develop control of attractors and their stability. 

In summary, the work extends the understanding of the generation of complex dynamics in a unidirectionally coupled universal oscillator, providing insights into a plethora of states bifurcating from \textit{isola}, quasiperiodic synchronization to chaos. 

\section*{Acknowledgements}
AP thanks Somrita Ray (IISER, Berhampur) for various supports. AP and SS acknowledge Rohitashwa Chattopadhyay (TIFR, Bombay) for some insightful discussions and suggestions.

\bibliographystyle{unsrt}
\bibliography{unicoupSLdeter.bib}

\begin{thebibliography}{10}

\bibitem{landau1944problem}
Lev~D Landau.
\newblock On the problem of turbulence.
\newblock In {\em Dokl. Akad. Nauk USSR}, volume~44, page 311, 1944.

\bibitem{stuart1958non}
John~Trevor Stuart.
\newblock On the non-linear mechanics of hydrodynamic stability.
\newblock {\em Journal of Fluid Mechanics}, 4(1):1--21, 1958.

\bibitem{stuart1960non}
John~Trevor Stuart.
\newblock On the non-linear mechanics of wave disturbances in stable and
  unstable parallel flows part 1. the basic behaviour in plane poiseuille flow.
\newblock {\em Journal of Fluid Mechanics}, 9(3):353--370, 1960.

\bibitem{kuramoto2003chemical}
Yoshiki Kuramoto.
\newblock {\em Chemical oscillations, waves, and turbulence}.
\newblock Courier Corporation, 2003.

\bibitem{zou2014emergence}
Wei Zou, DV~Senthilkumar, Jinqiao Duan, and J{\"u}rgen Kurths.
\newblock Emergence of amplitude and oscillation death in identical coupled
  oscillators.
\newblock {\em Physical Review E}, 90(3):032906, 2014.

\bibitem{roopnarain2021amplitude}
Ryan Roopnarain and S~Roy Choudhury.
\newblock Amplitude death, oscillation death, and periodic regimes in
  dynamically coupled landau--stuart oscillators with and without distributed
  delay.
\newblock {\em Mathematics and Computers in Simulation}, 187:30--50, 2021.

\bibitem{mendola2025collective}
Naveen~Kumar Mendola and Umeshkanta~Singh Thounaojam.
\newblock Collective dynamics and phase transitions of stuart-landau
  oscillators on a ring network: Interplay of asymmetric and symmetric
  couplings.
\newblock {\em Physical Review E}, 111(5):054218, 2025.

\bibitem{verma2025explosive}
Umesh~Kumar Verma.
\newblock Explosive transition in adaptive stuart-landau oscillators with
  higher-order interactions.
\newblock {\em Physical Review E}, 111(1):014302, 2025.

\bibitem{bayani2023explosive}
Atiyeh Bayani, Sajad Jafari, Hamed Azarnoush, Fahimeh Nazarimehr, Stefano
  Boccaletti, and Matja{\v{z}} Perc.
\newblock Explosive synchronization dependence on initial conditions: The
  minimal kuramoto model.
\newblock {\em Chaos, Solitons \& Fractals}, 169:113243, 2023.

\bibitem{abrams2004chimera}
Daniel~M Abrams and Steven~H Strogatz.
\newblock Chimera states for coupled oscillators.
\newblock {\em Physical Review Letters}, 93(17):174102, 2004.

\bibitem{parastesh2021chimeras}
Fatemeh Parastesh, Sajad Jafari, Hamed Azarnoush, Zahra Shahriari, Zhen Wang,
  Stefano Boccaletti, and Matja{\v{z}} Perc.
\newblock Chimeras.
\newblock {\em Physics Reports}, 898:1--114, 2021.

\bibitem{majhi2019chimera}
Soumen Majhi, Bidesh~K Bera, Dibakar Ghosh, and Matja{\v{z}} Perc.
\newblock Chimera states in neuronal networks: A review.
\newblock {\em Physics of life reviews}, 28:100--121, 2019.

\bibitem{majhi2024dynamical}
Soumen Majhi, Biswambhar Rakshit, Amit Sharma, J{\"u}rgen Kurths, and Dibakar
  Ghosh.
\newblock Dynamical robustness of network of oscillators.
\newblock {\em Physics Reports}, 1082:1--46, 2024.

\bibitem{garcia2012complex}
Vladimir Garc{\'\i}a-Morales and Katharina Krischer.
\newblock The complex ginzburg--landau equation: an introduction.
\newblock {\em Contemporary Physics}, 53(2):79--95, 2012.

\bibitem{lee2022nontrivial}
Seungjae Lee and Katharina Krischer.
\newblock Nontrivial twisted states in nonlocally coupled stuart-landau
  oscillators.
\newblock {\em Physical Review E}, 106(4):044210, 2022.

\bibitem{kemeth2019cluster}
Felix~P Kemeth, Sindre~W Haugland, and Katharina Krischer.
\newblock Cluster singularity: The unfolding of clustering behavior in globally
  coupled stuart-landau oscillators.
\newblock {\em Chaos: An Interdisciplinary Journal of Nonlinear Science},
  29(2):023107, 2019.

\bibitem{pikovsky2015dynamics}
Arkady Pikovsky and Michael Rosenblum.
\newblock Dynamics of globally coupled oscillators: Progress and perspectives.
\newblock {\em Chaos: An Interdisciplinary Journal of Nonlinear Science},
  25(9):097616, 2015.

\bibitem{duvsek1994numerical}
Jan Du{\v{s}}ek, Patrice Le~Gal, and Philippe Frauni{\'e}.
\newblock A numerical and theoretical study of the first hopf bifurcation in a
  cylinder wake.
\newblock {\em Journal of Fluid Mechanics}, 264:59--80, 1994.

\bibitem{kumar2021two}
Mohit Kumar and Michael Rosenblum.
\newblock Two mechanisms of remote synchronization in a chain of stuart-landau
  oscillators.
\newblock {\em Physical Review E}, 104(5):054202, 2021.

\bibitem{zhao2018enhancing}
Nannan Zhao, Zhongkui Sun, and Wei Xu.
\newblock Enhancing coherence via tuning coupling range in nonlocally coupled
  stuart--landau oscillators.
\newblock {\em Scientific Reports}, 8(1):8721, 2018.

\bibitem{haugland2023coexistence}
Sindre~W Haugland.
\newblock {\em On Coexistence Patterns: Hierarchies of Intricate Partially
  Symmetric Solutions to Stuart-Landau Oscillators with Nonlinear Global
  Coupling}.
\newblock Springer Nature, 2023.

\bibitem{li2022mean}
Yang Li, Jifan Shi, and Kazuyuki Aihara.
\newblock Mean-field analysis of stuart--landau oscillator networks with
  symmetric coupling and dynamical noise.
\newblock {\em Chaos: An Interdisciplinary Journal of Nonlinear Science},
  32(6):063114, 2022.

\bibitem{pandey2020coupled}
Ankan Pandey, A~Ghose-Choudhury, Partha Guha, et~al.
\newblock On coupled delayed van der pol-duffing oscillators.
\newblock {\em Journal of Applied Nonlinear Dynamics}, 9(04):567--574, 2020.

\bibitem{hashemi2025inter}
Parisa Hashemi, Fahimeh Nazarimehr, Fatemeh Parastesh, Farzad Towhidkhah, and
  Sajad Jafari.
\newblock Inter-layer coupling effects on synchronization in multiplex neuronal
  networks with different topologies.
\newblock {\em Chaos, Solitons \& Fractals}, 201:117205, 2025.

\bibitem{serrano2025fractional}
Fernando~E Serrano and Viet-Thanh Pham.
\newblock Fractional memristive-discrete neural network: projective terminal
  sliding mode synchronization.
\newblock {\em The European Physical Journal Special Topics}, pages 1--14,
  2025.

\bibitem{pham2019simulation}
Viet-Thanh Pham, Sajad Jafari, Christos Volos, and Luigi Fortuna.
\newblock Simulation and experimental implementation of a line--equilibrium
  system without linear term.
\newblock {\em Chaos, Solitons \& Fractals}, 120:213--221, 2019.

\bibitem{serrano2022robust}
Fernando~E Serrano and Dibakar Ghosh.
\newblock Robust stabilization and synchronization in a network of chaotic
  systems with time-varying delays.
\newblock {\em Chaos, Solitons \& Fractals}, 159:112134, 2022.

\bibitem{khennaoui2019chaos}
Amina-Aicha Khennaoui, Adel Ouannas, Samir Bendoukha, Giuseppe Grassi, Xiong
  Wang, Viet-Thanh Pham, and Fawaz~E Alsaadi.
\newblock Chaos, control, and synchronization in some fractional-order
  difference equations.
\newblock {\em Advances in Difference Equations}, 2019(1):412, 2019.

\bibitem{ghosh2022synchronized}
Dibakar Ghosh, Mattia Frasca, Alessandro Rizzo, Soumen Majhi, Sarbendu Rakshit,
  Karin Alfaro-Bittner, and Stefano Boccaletti.
\newblock The synchronized dynamics of time-varying networks.
\newblock {\em Physics Reports}, 949:1--63, 2022.

\bibitem{perlikowski2010routes}
Przemyslaw Perlikowski, Serhiy Yanchuk, Matthias Wolfrum, Andrzej Stefanski,
  Przemyslaw Mosiolek, and Tomasz Kapitaniak.
\newblock Routes to complex dynamics in a ring of unidirectionally coupled
  systems.
\newblock {\em Chaos: An Interdisciplinary Journal of Nonlinear Science},
  20(1):013111, 2010.

\bibitem{zhao2025nonreciprocal}
Luekai Zhao, Bojun Li, and Nariya Uchida.
\newblock Nonreciprocal amplification toward chaos in a chain of duffing
  oscillators.
\newblock {\em Physical Review E}, 112(1):014216, 2025.

\bibitem{barba2023dynamics}
JJ~Barba-Franco, A~Gallegos, R~Jaimes-Re{\'a}tegui, J~Mu{\~n}oz-Maciel, and
  AN~Pisarchik.
\newblock Dynamics of coexisting rotating waves in unidirectional rings of
  bistable duffing oscillators.
\newblock {\em Chaos: An Interdisciplinary Journal of Nonlinear Science},
  33(7):073126, 2023.

\bibitem{vijayan2025lag}
Vijeesh Vijayan, Hayder Natiq, Shaher Momani, Viet-Thanh Pham, and Matja{\v{z}}
  Perc.
\newblock Lag synchronization in an unidirectional ring of memristive neurons.
\newblock {\em The European Physical Journal Special Topics}, pages 1--12,
  2025.

\bibitem{ryu2017amplitude}
Jung-Wan Ryu, Jong-Ho Kim, Woo-Sik Son, and Dong-Uk Hwang.
\newblock Amplitude death in a ring of nonidentical nonlinear oscillators with
  unidirectional coupling.
\newblock {\em Chaos: An Interdisciplinary Journal of Nonlinear Science},
  27(8):083119, 2017.

\bibitem{pikovsky2001synchronization}
Arkady Pikovsky, Michael Rosenblum, J{\"u}rgen Kurths, and A~Synchronization.
\newblock A universal concept in nonlinear sciences.
\newblock {\em Self}, 2(3):10--1017, 2001.

\bibitem{nakao2016phase}
Hiroya Nakao.
\newblock Phase reduction approach to synchronisation of nonlinear oscillators.
\newblock {\em Contemporary Physics}, 57(2):188--214, 2016.

\bibitem{pyragas1992continuous}
Kestutis Pyragas.
\newblock Continuous control of chaos by self-controlling feedback.
\newblock {\em Physics Letters A}, 170(6):421--428, 1992.

\bibitem{Calogero2008}
Francesco Calogero.
\newblock {\em Isochronous systems}.
\newblock Oxford University Press, 2008.

\bibitem{limiso}
Sandip Saha and Gautam Gangopadhyay.
\newblock Isochronicity and limit cycle oscillation in chemical systems.
\newblock {\em Journal of Mathematical Chemistry}, 55(3):887--910, Mar 2017.

\bibitem{calogero2011isochronous}
Francesco Calogero.
\newblock Isochronous dynamical systems.
\newblock {\em Philosophical Transactions of the Royal Society A: Mathematical,
  Physical and Engineering Sciences}, 369(1939):1118--1136, 2011.

\bibitem{saha2025power}
Sandip Saha.
\newblock Power law behavior of center-like decaying oscillation: Exponent
  through perturbation theory and optimization.
\newblock {\em Communications in Nonlinear Science and Numerical Simulation},
  148:108844, 2025.

\bibitem{saha2022existence}
Sandip Saha and Gautam Gangopadhyay.
\newblock The existence of a stable limit cycle in the
  li{\'e}nard--levinson--smith (lls) equation beyond the lls theorem.
\newblock {\em Communications in Nonlinear Science and Numerical Simulation},
  109:106311, 2022.

\bibitem{koseska2013oscillation}
Aneta Koseska, Evgeny Volkov, and J{\"u}rgen Kurths.
\newblock Oscillation quenching mechanisms: Amplitude vs. oscillation death.
\newblock {\em Physics Reports}, 531(4):173--199, 2013.

\bibitem{rosenblum2004controlling}
Michael~G Rosenblum and Arkady~S Pikovsky.
\newblock Controlling synchronization in an ensemble of globally coupled
  oscillators.
\newblock {\em Physical Review Letters}, 92(11):114102, 2004.

\bibitem{reddy1998time}
DV~Ramana Reddy, Abhijit Sen, and George~L Johnston.
\newblock Time delay induced death in coupled limit cycle oscillators.
\newblock {\em Physical Review Letters}, 80(23):5109, 1998.

\end{thebibliography}
\end{document}